\documentclass[aps, superscriptaddress,nofootinbib, twocolumn, notitlepage]{revtex4-1}
\usepackage{amsmath,amsfonts, amssymb, amsthm, dsfont}
\usepackage{yfonts}
\usepackage{bm}
\usepackage{mathrsfs}
\usepackage{graphicx}
\usepackage{verbatim}
\usepackage{hyperref}
\usepackage{wasysym}
\usepackage{pst-all}
\usepackage{leftidx}

\usepackage{tikz,pgfplots}\pgfplotsset{compat=newest}
\usetikzlibrary{external,calc,decorations.pathreplacing,decorations.markings,decorations.pathmorphing,arrows.meta,shapes.geometric}

\newcommand{\di}{\mathrm{d}}

\newcommand{\ket}[1]{|#1\rangle}

\renewcommand{\ol}[1]{\overline{#1}}
\newcommand{\comments}[1]{}
\newcommand{\mb}[1]{\mathbf{#1}}
\renewcommand{\cal}[1]{\mathcal{#1}}

\newcommand{\coho}[1]{\textswab{#1}}
\newcommand{\cohosub}[1]{\scalebox{0.7}{\textswab{#1}}}

\newcommand{\ag}[2]{#1_\mb{#2}}

\def\U{\mathrm{U}(1)}
\def\H{\mathcal{H}}
\def\Z{\mathbb{Z}}

\def\Aut{\mathrm{Aut}}

\DeclareMathOperator{\tr}{Tr}

\makeatletter
\def\l@subsubsection#1#2{}
\makeatother

\begin{document}

\title{Relative Anomaly in (1+1)d Rational Conformal Field Theory}

\author{Meng Cheng}
\affiliation{Department of Physics, Yale University, New Haven, CT 06511-8499, USA}
\author{Dominic J. Williamson}
\affiliation{Department of Physics, Yale University, New Haven, CT 06511-8499, USA}
\affiliation{Stanford Institute for Theoretical Physics, Stanford University, Stanford, CA 94305}
\date{\today}
\begin{abstract}
 We study 't Hooft anomalies of symmetry-enriched rational conformal field theories (RCFT) in (1+1)d. Such anomalies determine whether a theory can be realized in a truly (1+1)d system with on-site symmetry, or on the edge of a (2+1)d symmetry-protected topological phase. 
 RCFTs with the identical symmetry actions on their chiral algebras may have different 't Hooft anomalies due to additional symmetry charges on local primary operators.  To compute the relative anomaly, we establish a precise correspondence between (1+1)d non-chiral RCFTs and (2+1)d doubled symmetry-enriched topological (SET) phases with a choice of symmetric gapped boundary.  Based on these results we derive a general formula for the relative 't Hooft anomaly in terms of algebraic data that characterizes the SET phase and its boundary. 
\end{abstract}
\maketitle

\tableofcontents

\section{Introduction}
Boundary states of a symmetry-protected topological (SPT) phase~\cite{chen2013} exhibit a 't Hooft anomaly, and thus can not be trivially gapped, i.e. can not form symmetric short-range entangled (SRE) states. For a (2+1)d SPT phase, the (1+1)d edge is either gapless or symmetry-breaking. Symmetry-protected gapless edge states are often taken as a hallmark of topologically nontrivial bulk phases. 

Often, such edge states can be described by conformal field theories (CFT) at low energy. For example, many bosonic SPT phases have a Luttinger liquid edge theory~\cite{lu2012, LevinGu, NingSPTedge}.
Given a CFT with a global symmetry group $G$, a natural question is whether this ``symmetry-enriched'' CFT must be realized on the boundary of a (2+1)d SPT phase, i.e. whether the theory has a 't Hooft anomaly. In principle, the anomaly can be computed by studying correlation functions of symmetry defect operators. In practice, this may not be easy to implement, and the connection to the local operator algebra remains obscure. A number of alternative ways to compute the 't Hooft anomaly in a CFT have been developed, and a common method is to consider the modular covariance properties of symmetry-twisted partition functions on general Riemann surfaces~\cite{FelderCMP1988, FreedVafa1987, RyuZhang, SulePRB2013, hsieh2014}, which signals the obstruction to orbifolding. This method works well for cyclic groups such as $\Z_n$, whose anomalies can be detected by modular properties of torus partition functions, but can become cumbersome for more general groups, in particular when the anomaly can only be seen on higher-genus surfaces~\cite{Mignard2017, Bonderson2018, WenXD2019}.

In this work, we are mainly concerned with the computation of ``relative'' anomalies for rational CFTs (RCFT) enriched by unitary symmetry. (1+1)d RCFTs are defined by having a finite number of primary operators. A remarkable property of any RCFT is that its chiral/holomorphic half can be regarded as the edge theory of a (2+1)d chiral topological phase~\cite{witten1989, Moore1991, wen1990edge}, whose anyon excitations are in one-to-one correspondence with chiral primary operators. The full non-chiral theory can be obtained by ``gluing'' the chiral and the anti-chiral parts together consistently, such that the partition function on an arbitrary Riemann surface is modular-invariant ( i.e. invariant under the action of the mapping class group). More precisely, the full RCFT corresponds to a \emph{doubled} topological phase with a gapped boundary~\cite{FRS1}. This bulk-boundary correspondence presents a new way to understand 't Hooft anomalies in symmetry-enriched RCFTs. 

Generally, symmetry-enriched CFTs are characterized by symmetry actions on local scaling operators. For (1+1)d RCFTs, a distinguished class of local operators are those that are fully chiral/anti-chiral, which form the extended chiral algebra of the RCFT. In essence, the chiral algebra describes the mathematical structure of local operators on the chiral edge of the corresponding (2+1)d bulk, and thus completely defines the topological order that underlies the chiral RCFT. When global symmetry is taken into account, we show that the symmetry action on the chiral algebra uniquely defines the symmetry-enriched topological order of the bulk.  Once this is given, different symmetry-enriched RCFTs can further differ by additional ``charges'' on local primary operators, which correspond to quantum numbers of anyons condensing on the gapped boundary. This charge assignment may change the 't Hooft anomaly of the theory, which is defined as the relative anomaly.

We relate relative 't Hooft anomalies of a (1+1)d RCFT to the question of symmetric gapped boundary conditions for a (2+1)d topological phase~\cite{Bischoff2019}, enriched by a global symmetry. 
We classify such boundary conditions and, in the case of doubled SET phases that are relevant to RCFT, use an exactly solvable lattice model to find a formula for the relative anomaly purely in terms of algebraic data that describes the (2+1)d bulk.  As a by-product, the same result also provides a sufficient condition for when a $(2+1)$d SET phase can ``absorb'' an SPT phase. 

\section{Symmetric gapped boundaries of doubled SET phases}
We first study symmetric gapped boundaries of a special family of (2+1)d SET phases, namely those that can be viewed as two separate layers with opposite SET order, i.e. ``doubled''. It turns out that this problem is closely related (in a way equivalent) to the classification of relative anomalies in (1+1)d RCFTs. We develop a general theory of such gapped boundaries within the mathematical framework of $G$-crossed braided tensor categories~\cite{ENO2009, SET}, which is reviewed briefly below. A brief summary of the basic notions of a modular tensor category (MTC), the algebraic theory of anyons in (2+1)d gapped phases, is provided in Appendix~\ref{sec:mtc-review}. 

\subsection{Review of the bulk classification}
First we briefly recall the classification of (2+1)d SET phases~\cite{SET, Tarantino_SET, Chen2014}, mainly following Ref.~\cite{SET}. Given a bulk MTC $\cal{C}$, a fundamental property is its ``topological symmetry group'' $\mathrm{Aut}(\cal{C})$, which consists of all permutations of anyon types preserving their universal topological properties.
 Global symmetry transformations can permute anyons, specified by a group homomorphism $\rho: G\rightarrow\mathrm{Aut}(\cal{C})$. 

Anyon excitations can also transform projectively under $G$, a phenomenon known as symmetry fractionalization. Let us discuss the simpler case where global symmetries do not permute anyon types. Let $R_\mb{g}$ be the representation of $\mb{g}\in G$ on the full Hilbert space. Consider an excited state with $n$ well-separated anyons $a_1,\cdots, a_n$. Since the ground state is symmetric, we expect that the $G$ action can be localized to the neighborhood of $a_j$, as a local unitary operator $U_\mb{g}^{(j)}$. More precisely, for any local operator $O$ supported on a small neighborhood of $a_j$, we should have
\begin{equation}
	R_\mb{g}OR_\mb{g}^{-1}\approx U_\mb{g}^{(j)}O(U_\mb{g}^{(j)})^{-1}.
	\label{}
\end{equation}
Thus, within the subspace of states with fixed anyons $a_1, a_2,\cdots, a_n$, we have the following decomposition:
\begin{equation}
	R_\mb{g}\approx\prod_j U_\mb{g}^{(j)}.
	\label{eqn:Rdecomp}
\end{equation}
We expect that this assumption, that the global symmetry transformation can be ``localized'' must be true for any on-site symmetries.

In general these local unitary operators only form projective representations of $G$:
\begin{equation}
	U_\mb{g}^{(a)}U_\mb{h}^{(a)}=\eta_a(\mb{g,h})U_\mb{gh}^{(a)},
	\label{}
\end{equation}
where $\eta_a\in \U$ is referred to as the projective phases (or factor set).  One can show that 
\begin{equation}
	\eta_a(\mb{g,h})\eta_b(\mb{g,h})=\eta_c(\mb{g,h}).
	\label{eqn:eta-fusion}
\end{equation}
whenever $N_{ab}^c>0$, which ensures that the physical states transform regularly under $G$. Thus one can express $\eta_a(\mb{g,h})$ as the braiding of $a$ with an Abelian anyon $\coho{w}(\mb{g,h})$:
\begin{equation}
	\eta_a(\mb{g,h})=M_{a, \cohosub{w}(\mb{g,h})}.
	\label{}
\end{equation}
It is shown in Ref.~\cite{SET} that $\coho{w}(\mb{g,h})$, called the fractionalization class, is classified by cohomology classes in $\H^2[G, \cal{A}]$, where $\cal{A}$ is the group of Abelian anyons in $\cal{C}$. The trivial class, where one can set $\eta_a(\mb{g,h})\equiv 1$ for all $a$, corresponds to an SET phase where the symmetry acts trivially.

This classification is generalized to symmetries permuting anyon types in Ref.~\cite{SET}, in other words the group homomorphism $\rho$ is nontrivial. The crucial difference is the following: while locally in the neighborhood of anyon excitations, $R_\mb{g}$ is still approximated by $U_\mb{g}^{(j)}$'s, globally the decomposition Eq.~\eqref{eqn:Rdecomp} fails, as one also needs to take into account the nontrivial action on the topological fusion space. Intuitively, the symmetry transformation also acts on the ``splitting'' operators that are used to create the state with multiple anyons, given by the so-called $U$ symbols (see for more details). Thus the relation Eq.~\eqref{eqn:eta-fusion} must be modified accordingly. 
It turns out that not every $\rho$ is compatible with symmetry localization. The failure is captured by an obstruction class in $\H^3_\rho[G,\mathcal{A}]$. If the obstruction class is nontrivial cohomologically, then $\rho$ can not be realized by an on-site symmetry group $G$ in a pure 2D system.

When the symmetry localization obstruction vanishes, one can show that $\H^2_\rho[G, \cal{A}]$ gives a ``torsor'' over all fractionalization classes. Namely, given $[\coho{w}]\in \H^2_\rho[G, \cal{A}]$, one can modify the projective phases
\begin{equation}
	\eta_a(\mb{g,h})\rightarrow \eta_a(\mb{g,h})M_{a,\cohosub{w}(\mb{g,h})}
	\label{}
\end{equation}
to arrive at a new fractionalization class, and hence a distinct SET phase.

When the symmetry is finite and unitary, extrinsic defects carrying symmetry fluxes can be introduced to the system to characterize the SET order. An algebraic theory for symmetry defects is formulated in Ref.~\cite{SET}, known mathematically as a $G$-crossed braided tensor category. The defect theory can be consistently defined if and only if an obstruction class in $\H^4[G, \U]$ vanishes. Otherwise, the SET must exist on the surface of a 3D $G$ bosonic SPT phase~\cite{Chen2014}. When the $\H^4$ obstruction class is trivial, the defect theory can be modified by a $3$-cocycle $[\alpha]$ in $\H^3[G, \U]$, physically corresponding to stacking 2D bosonic SPT phases. This stacking can only affect the properties of symmetry defects, and possibly edge excitations, but does not change anything about anyon excitations in the bulk. 

While each choice of $\rho$ and $[\coho{w}]$ yields a distinct SET phase, different choices of $[\alpha]$ may actually lead to the \emph{same} SET phase. In other words, an SET phase may ``trivialize'' an SPT phase. In the following we denote the SPT phase with $3$-cocycle $[\alpha]$ by $\mathrm{SPT}_{[\alpha]}$, and the stacking operation by $\boxtimes$.

\begin{figure}[t]
	\centering
	\includegraphics[width=\columnwidth]{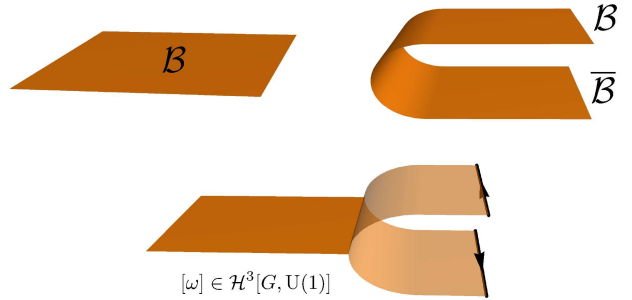}
	\caption{A symmetric gapped boundary between $\cal{B}\boxtimes \ol{\cal{B}}$ and SPT$_{[\omega]}$, equivalent to the trivialization of SPT$_{[\omega]}$ by $\cal{B}$. }
	\label{fig:folding}
\end{figure}

Suppose that an SET phase, denoted abstractly by $\cal{B}$, can trivialize an SPT phase with $[\alpha]\in \H^3[G, \U]$. This means that $\cal{B}$ and $\cal{B}\boxtimes \mathrm{SPT}_{[\alpha]}$ are in the same symmetry-enriched phase.
Since in the absence of symmetry defects the SPT phase merely changes boundary excitations, it should be possible to form a completely gapped and symmetric boundary between $\cal{B}\boxtimes \mathrm{SPT}_{[\alpha]}$ and $\cal{B}$. If we ``fold'' $\cal{B}$, i.e. take its conjugate theory $\ol{\cal{B}}$~\footnote{Here the conjugate of $\cal{B}$ can be defined algebraically, namely one defines another $G$-crossed braided category, with the same anyons and defects as $\cal{B}$, the same fusion rules, but all other data are complex conjugate of those of $\cal{B}$. The corresponding SET phase is $\ol{\cal{B}}$.} then we have a gapped and symmetric boundary between $\cal{B}\boxtimes\ol{\cal{B}}$ and $\mathrm{SPT}_{[\alpha]}$, as illustrated in Fig.~\ref{fig:folding}. Thus we conclude that the existence of a gapped and symmetric boundary between $\cal{B}\boxtimes\ol{\cal{B}}$ and $\mathrm{SPT}_{[\alpha]}$ is equivalent to the trivialization of SPT$_{[\alpha]}$ by the SET phase $\cal{B}$. In the following we will study symmetric gapped boundaries of $\cal{B}\boxtimes\ol{\cal{B}}$ in great detail.

\subsection{Gapped boundary of doubled SET phases}
\label{sec:SETboundary}
Before discussing the general theory, let us study an example of the interface between a $\Z_2$ SPT phase and a doubled $\U_{2n}$ topological phase (the $n=1$ case was considered in Ref.~\cite{LuPRB2014}).  Edge excitations of both phases can be described by multi-component Luttinger liquids~\cite{wengaplessboundary}, whose action generally takes the following form:
\begin{equation}
	S=\int\di t\di x\,\left(\frac{1}{4\pi}\partial_t\Phi^\mathsf{T}K\partial_x\Phi - \frac{1}{4\pi}\partial_x\Phi^\mathsf{T}V \partial_x\Phi+ \cdots\right).
	\label{}
\end{equation}
Here $\Phi$ denotes a column vector of bosonic fields, all of which are $2\pi$-periodic, and $K$ is a symmetric integer matrix that encodes the commutation relations between $\Phi$ fields.

The edge of the doubled $\U_{2n}$ state can be described using $K=\begin{pmatrix} 2n & 0\\ 0 & -2n \end{pmatrix}$, with the two fields denoted by $\varphi_1$ and $\varphi_2$. The SPT edge theory is a Luttinger liquid with $K=\begin{pmatrix} 0 & 1\\ 1 & 0\end{pmatrix}$, and the two fields are denoted by $\varphi$ and $\theta$. Under the $\Z_2$ symmetry they transform as~\cite{lu2012}
\begin{align}
	\varphi\rightarrow \varphi+\pi, && \theta\rightarrow \theta+\pi.
	\label{}
\end{align}
The $\Z_2$ transformations of $\varphi_1$ and $\varphi_2$ are given by
\begin{align}
	\varphi_1\rightarrow \varphi_1+\frac{\pi p}{2n}, &&  \varphi_2\rightarrow \varphi_2-\frac{\pi p}{2n}, 
	\label{}
\end{align}
where $p$ is an integer. We remark that $\varphi_1$ and $\varphi_2$ have opposite phase shifts since the two ``layers'' are supposed to be conjugate to each other. The reason that the phases are quantized in this way is to make sure local operators, in particular $e^{2n\pi i\varphi_{1,2}}$ only pick up $\pm 1$ phase factors under the symmetry. In fact, $p$ being even (odd) corresponds to exactly the trivial (nontrivial) fractionalization class of $\Z_2$ in $\U_{2n}$, since $\H^2[\Z_2, \Z_{2n}]=\Z_2$.

The SPT and the doubled $\U_{2n}$ edges are coupled through the following gapping terms:
\begin{equation}
	{L}=\Delta\big[\cos (2n\varphi_1+n\varphi-\theta)+\cos(2n\varphi_2+n\varphi+\theta)\big].
	\label{}
\end{equation}
On this edge, $\varphi+\varphi_1+\varphi_2$ is pinned (when $n=1$, $\varphi_1-\varphi_2-\theta$ is also pinned). Demanding both terms preserve the global $\Z_2$ symmetry, we have
\begin{equation}
	p+n+1\equiv 0 \,(\text{mod }2).
	\label{}
\end{equation}
One can show that the edge is gapped out without local degeneracy.
We remark that for odd $n$, $n+1$ is already even so $p$ can be set to $0$. However, for even $n$ it is necessary to take $p=1$. In other words, the $\Z_2$ symmetry must be fractionalized~\cite{LevinPRB2012, lu2013} for even $n$.

Now let us consider the other edge of the doubled $\U_{2n}$ state, denoted by $\varphi_1'$ and $\varphi_2'$. As edges of the same layer, we may assume that $\varphi_a$ and $\varphi_a'$ are in fact the ``same'' edge modes. This means vertex operators $e^{i\varphi_a}$ and $e^{i\varphi_a'}$ are identified with the same bulk anyon type. In other words, $e^{i(\varphi_a-\varphi_a')}$ tunnels an anyon from one edge to another. The $K$ matrix for the primed fields is $\begin{pmatrix} -2n & 0\\ 0 & 2n \end{pmatrix}$, in the basis of $\varphi_1', \varphi_2'$. The $\Z_2$ symmetry transformation takes the same form:
\begin{align}
	\varphi_1'\rightarrow \varphi_1'+\frac{\pi p}{2n}, && \varphi_2'\rightarrow \varphi_2'-\frac{\pi p}{2n}.
	\label{}
\end{align}
On this edge, one can add a gapping term
\begin{equation}
	L'\sim \cos 2n(\varphi_1'+\varphi_2'),
	\label{}
\end{equation}
which gaps out all edge modes without breaking the $\Z_2$ symmetry. Thus we have a symmetric gapped boundary between the doubled $\U_{2n}$ SET phase and the vacuum.

Now if we consider a strip of the doubled $\U_{2n}$ SET phase, between the $\Z_2$ SPT phase and the vacuum, where both boundaries are gapped and symmetric (using $L$ and $L'$, respectively). Naively it may seem that one has constructed a symmetric gapped boundary between the $\Z_2$ SPT phase and the vacuum, as there is no local symmetry-breaking order on either edges. This kind of ``paradox'' was studied in Ref.~\cite{WangPRB2013} and the resolution is the following: the symmetry is broken by a string operator connecting the two edges.
\begin{equation}
	W=e^{i(\varphi_1'-\varphi_1-\varphi_2+\varphi_2'-\varphi)}.
	\label{}
\end{equation}
It is easy to see that $W$ has a finite expectation value, and under the $\Z_2$ symmetry $W\rightarrow -W$. Effectively, if one decreases the width of the SET strip to make it quasi-1D, $W$ becomes a local order parameter on a gapped boundary which spontaneously breaks the symmetry.

We now move on to a more abstract but general description of gapped boundaries for 2D topological phases.
The theory can be formulated in a number of different ways. We will describe the boundary in terms of anyon condensation~\cite{LevinPRX2013,WangPRB2015, lan2015, kong2014}. Another description is based on the fact that a topological phase which admits a fully gapped boundary must be a quantum double of a unitary fusion category and can be realized by a generalized string-net model. Then one may directly construct and classify gapped boundaries in the lattice model~\cite{kitaev2012, HuJHEP2018}. This approach is used for the full calculation in Appendix \ref{sec:set-model}.

Suppose the bulk is described by a MTC $\cal{C}$. A gapped boundary corresponds to a Lagrangian algebra $\cal{L}$ of the bulk MTC $\cal{C}$. Mathematically a Lagrangian algebra is a general object
\begin{equation}
	\cal{L}=\sum_{a\in \cal{C}} n_a a.
	\label{}
\end{equation}
Here $n_a\geq 0$ are non-negative integers, satisfying $\sum_a n_a d_a=\mathcal{D}$, where $\mathcal{D}=\sqrt{\sum_{a\in\mathcal{C}}d_a^2}$ is the total quantum dimension of $\cal{C}$.  Physically, if $n_a>0$, it means that the anyon $a$ is condensed on the boundary. It follows that such condensed anyons must be bosonic and have trivial mutual braiding statistics with each other. There are other conditions, such as associativity condition on the fusion of anyons in the algebra, to be satisfied for the anyons to condense and we refer to Refs.~\cite{kong2014, eliens2013, Cong2017} for more comprehensive account of the theory. In the previous example of doubled $\U_{2n}$ theory, the pinned field $\varphi_1+\varphi_2+\varphi$ exactly corresponds to a condensed bosonic anyon.

Here we specialize to the case where $\cal{C}=\cal{B}\boxtimes\ol{\cal{B}}$ for some MTC $\cal{B}$ that corresponds to the SET phase. We denote the anyons in $\cal{C}$ by $(a,a')$ where $a,a'\in \cal{B}$. The most natural boundary condition corresponds to an anyon from one layer becoming the same type of anyon on the other layer via the folding. Equivalently, $(a,a)$ condenses on the boundary. Thus the Lagrangian algebra is given by
\begin{equation}
	\cal{L}=\sum_{a\in \cal{B}}(a,a).
	\label{eqn:L-diag}
\end{equation}
Alternatively, when anyons go from one layer to the other they may be transformed by an automorphism $\varphi\in \mathrm{Aut}(\cal{B})$, which corresponds to a Lagrangian algebra
\begin{equation}
	\cal{L}=\sum_{a\in \cal{B}}\big(a,\varphi(a)\big).
	\label{eqn:lag1}
\end{equation}
This is equivalent to folding along a nontrivial invertible domain wall. We note that there may be other Lagrangian algebras in $\cal{C}$ different from those given in Eq. \eqref{eqn:lag1}, for example when anyons in $\cal{B}$ or $\ol{\cal{B}}$ alone condense, corresponding to folding along a non-invertible domain wall.

Let us spell out the Lagrangian algebra for the doubled $\U_{2n}$ theory. Denote anyon types in a (single) $\U_{2n}$ theory by $[j], j=0,1,\cdots, 2n-1$. Then the Lagrangian algebra is $\cal{L}=\sum_{j=0}^{2n-1}(j,j)$.
As we have already mentioned, the $(1,1)$ boson, which corresponds to the field $e^{i(\varphi_1+\varphi_2+\varphi)}$ in the edge theory, condenses on the boundary.

Now we take into account the global symmetry and consider gapped boundaries preserving the global symmetry group. Intuitively,  in order to have such a symmetric gapped boundary it is reasonable to impose the following genereal conditions on the condensed anyons:
\begin{enumerate}
	\item The Lagrangian algebra must be invariant, i.e. $\rho_\mb{g}(\cal{L})=\cal{L}$ for each $\mb{g}\in G$. Therefore, if $a$ is condensed, $\rho_\mb{g}(a)$ must be in the condensation as well. 
	\item Condensed anyons can not carry any projective quantum numbers, or multi-dimensional representations.
\end{enumerate}
The first condition is fairly obvious, as if $\cal{L}$ is not invariant, then the gapped boundary explicitly breaks the symmetry. 
The second condition essentially ensures that there is no spontaneous symmetry breaking on the boundary. We give a more precise definition in Appendix \ref{sec:charged-condensate}.  

%To see why this is the case, consider $\rho=\mathds{1}$, so the symmetry fractionalization is classified by $\H^2[G, \cal{A}]$. On an anyon $a$, there are two types of projective representations: in one case, the factor set $\{\eta_a(\mb{g,h})\}$ defines a nontrivial $2$-cocycle in $\H^2[G, \U]$, and thus the anyon carries a protected local degeneracy. In the other case, while $\{\eta_a(\mb{g,h})\}$ belongs to the trivial class in $\H^2[G, \U]$, $a$ may still carry fractional charges of $G$. 
%\begin{equation}
%	\eta_a=\delta \lambda_a, 
%	\label{}
%\end{equation}
%Define $q_{a,b;c}(\mb{g})=\lambda_a(\mb{g})\lambda_b(\mb{g})\lambda_c^{-1}(\mb{g})$ for all $N_{ab}^c>0$. Clearly $\delta q_{a,b;c}=1$, so $q_{a,b;c}$ is a one-dimensional representation of $G$. If the fractionalization class is nontrivial, there exists a triplet $a,b,c$ such that $q_{a,b;c}(\mb{g})\neq 1$ for some $\mb{g}\in G$. Therefore, the local splitting operator that transforms $c$ to $a$ and $b$ is charged under $\mb{g}$. If $a,b,c$ all condense, the local splitting operator preserves the condensation as well and thus the $G$ symmetry is spontaneously broken.  

Therefore, the only remaining freedom is to have condensed anyons being charged under the symmetry group. Here by ``charge'' we mean one-dimensional representations. Let us classify such charged condensation for the doubled SET phase $\cal{Z}(\cal{B})$, with the Lagrangian algebra $\mathcal{L}=\sum_{a\in \mathcal{B}}\big(a,\varphi(a)\big)$. Suppose that $\rho$ is trivial for simplicity. Since the SET order in the two layers are conjugate to each other, $(a,a)$ carries no projective representation. Denote the one-dimensional representation carried by $\big(a,\varphi(a)\big)$ by $\phi_a$. These representations should satisfy fusion rules: since $(c,c)$ appears in the fusion channels of $(a,a)$ and $(b,b)$ if $N_{ab}^c>0$:
\begin{equation}
	\phi_a(\mb{g})\phi_b(\mb{g})=\phi_c(\mb{g}), \quad N_{ab}^c>0.
	\label{}
\end{equation}
Thus we can express
\begin{equation}
	\phi_a(\mb{g})=M_{a, \cohosub{v}(\mb{g})}^*,
	\label{}
\end{equation}
where $\coho{v}(\mb{g})\in \cal{A}$. The requirement that $\phi_a(\mb{g})$ forms a representation means $\coho{v}(\mb{g})\times\coho{v}(\mb{h})=\coho{v}(\mb{gh})$, i.e. $\coho{v}$ belongs to $\H^1[G, \cal{A}]$. 
%We also notice that because a one-dimensional representation is necessarily a class function, so is $\coho{v}$. We will sometimes write $\coho{v}([\mb{g}])$ to emphasize this property. The identity representation will be denoted by $\mathds{1}$.

This discussion can be generalized to the case where anyons are permuted by the symmetry. A general definition of symmetry-preserving anyon condensation is presented in Appendix \ref{sec:charged-condensate}. In this case, one finds that $[\coho{v}]\in\H^1_\rho[G, \cal{A}]$ is only a \emph{torsor} over different types of charged condensations. 

We now determine whether this is a gapped boundary between the doubled SET phase and an SPT phase. Notice that there is a ``canonical'' gapped boundary between the doubled SET phase and the vacuum from the folding construction. It corresponds to all the condensed anyons being neutral under the symmetry. Suppose we modify the charges with $[\coho{v}]\in\H^1_\rho[G, \cal{A}]$. One would like to know what is the resulting phase after condensing the Lagrangian algebra.

To this end, it is useful to gauge the $G$ symmetry. Namely, the SET phase is coupled to a dynamical $G$ gauge field. Roughly speaking, gauging introduces $G$ fluxes to the theory, and projects to the gauge-invariant subspace. In particular, an anyon in the original theory can transform under different representations of the gauge group, which all become topologically distinct excitations after gauging. Thus the Lagrangian algebra becomes a condensable algebra in the resulting MTC. Condensing this algebra should leave behind a pure $G$ gauge theory, as all the anyons from the SET phase become confined.

Generally gauging can be quite complicated, so let us consider a simplified case where the symmetry acts trivially on anyons (i.e. $\rho\equiv\mb{1}$ and $[\coho{w}]=[0]$). In this case, the gauged MTC can be written as $\cal{Z}(\cal{B})\boxtimes \mathrm{D}(G)$. Here $\mathrm{D}(G)$ represents the topological order of an (untwisted) $G$ gauge theory. Excitations in $\mathrm{D}(G)$ can be labeled by their gauge flux, i.e. a conjugacy class $[\mb{g}]$ where $\mb{g}$ is a representative element, and an irreducible representation $\pi$ of the centralizer group of $\mb{g}$. We thus label a general anyon in the fully gauged theory as $\big( (a,a'), [\mb{g}], \pi\big)$. The Lagrangian algebra naturally lifts to the following algebra
\begin{equation}
	\cal{A}_G=\sum_{a\in\cal{B}}\big((a,\varphi(a)), [1], \phi_a\big).
	\label{}
\end{equation}
Condensing $\cal{A}_G$ confines all anyons in $\cal{Z}(\cal{B})$.  In fact, it is easy to check that only anyons of the following form survive the condensation:
\begin{equation}
	\begin{split}
	\big( &(a\times \coho{v}([\mb{g}]),\varphi(a)), [\mb{g}], \pi\big) \\
	&=\big( (\coho{v}([\mb{g}]),1), [\mb{g}], \pi\otimes \ol{\phi}_a\big)\times\big((a,\varphi(a)), [1], \phi_a\big).
\end{split}
	\label{}
\end{equation}
Here we make it explicit that a one-dimensional representation must be a class function. Thus the remaining anyons are given by $\big( (\coho{v}([\mb{g}]),1), [\mb{g}], \pi)$. As all gauge charges remain deconfined, the resulting phase is still a $G$ gauge theory, but now twisted by the following 3-cocycle:
\begin{equation}
	\omega(\mb{g,h,k})=F^{\cohosub{v}(\mb{g}), \cohosub{v}(\mb{h}), \cohosub{v}(\mb{k})}.
	\label{eqn:anomaly1}
\end{equation}

When the bulk fractionalization class is nontrivial, we obtain a general expression for the 3-cocycle with the diagonal condensation in Eq. \eqref{eqn:L-diag}:
\begin{equation}
	\begin{split}
	\omega(\mb{g,h,k})=& F^{\cohosub{v}(\mb{g}), {}^\mb{g}\cohosub{v}(\mb{h}), {}^\mb{gh}\cohosub{v}(\mb{k})}\\
	&\cdot U_\mb{g}^{-1}\big({}^\mb{g}\coho{v}(\mb{h}), {}^\mb{gh}\coho{v}(\mb{k})\big) \eta_{ {}^\mb{gh}\cohosub{v}(\mb{k})}(\mb{g,h})
	\end{split}
	\label{eqn:anomaly2}
\end{equation}
Here ${}^\mb{g}x\equiv \rho_\mb{g}(x)$. The formula is written in terms of $U$ and $\eta$ symbols, defined in Appendix \ref{sec:mtc-review}. The derivation, based on a symmetry-enriched string-net construction of the doubled SET phase, can be found in Appendix \ref{sec:set-model}.

When $\rho\equiv \mathds{1}$, one makes a canonical choice $U_\mb{g}(a,b;c)=1$, and $\eta_{a}(\mb{g,h})=M_{\cohosub{w}(\mb{g,h}),a}$. The anomaly formula becomes
\begin{equation}
	\omega(\mb{g,h,k})=F^{\cohosub{v}(\mb{g}), \cohosub{v}(\mb{h}), \cohosub{v}(\mb{k})}M_{\cohosub{w}(\mb{g,h}), \cohosub{v}(\mb{k}) }.
	\label{eqn:anomaly3}
\end{equation}

Let us explain the formula in more physical albeit heuristic terms. From the gauging construction above, one can see that due to the symmetry charges of the condensed anyons, gauge fluxes must be ``dressed'' with additional Abelian anyons $\coho{v}$'s to be deconfined after condensation. The two terms in Eq. \eqref{eqn:anomaly3} can now be understood easily, as the $F$ symbols of the dressed Abelian anyons and the projective representations that defects now carry due to the dressing.  At the level of defect theory (i.e. $G$-crossed braided tensor category), the dressing amounts to a relabeling of $\mb{g}$ defects by $a_\mb{g}\rightarrow a_\mb{g}\times \coho{v}(\mb{g})$ for each $a_\mb{g}\in \cal{C}_\mb{g}$. Such a relabeling does not affect the symmetry action on anyons, but can result in a change of defect properties. From the general classification, the effect is equivalent to the stacking of an SPT phase. Related results have been obtained in Ref. \cite{Hsin2019} from the point of view of higher-form symmetries.

This example demonstrates a general phenomenon: for a symmetric Lagrangian algebra to create a symmetric gapped boundary to vacuum via its condensation, an obstruction valued in $\H^3[G,\U]$ must vanish. Physically this obstruction means the condensation leads to an SPT phase~\cite{JiangPRB2017, LeePRB2018}.

\section{Review of rational conformal field theories}
Here we provide a brief review of RCFTs~\cite{difrancesco}.
In (1+1)d CFT, left-moving and right-moving conformal symmetries are decoupled (under e.g. periodic boundary conditions).  The (holomorphic) energy-momentum tensor $T(z)$, together with other mutually commuting holomorphic operators, such as conserved currents, form the chiral algebra $\cal{V}$ of the CFT. Mode expansions of these operators generate infinite-dimensional algebras.  Any CFT has the Virasoro algebra from stress tensor, and one often encounters affine Kac-Moody algebra from Lie group symmetry, e.g. Wess-Zumino-Witten (WZW) models. Higher-spin extensions of the Virasoro algebra, generally known as $\cal{W}$-algebras, can also occur, e.g. in coset constructions~\footnote{In this work we only allow integer-spin operators in the chiral algebra, thus excluding superconformal algebras.}. The chiral algebra characterizes the structure of holomorphic local scaling operators. We write $\mathcal{V}_L$ to emphasize that this is the algebra of chiral (left-moving) operators. In the full CFT, the identity sector (i.e. all descendants of the identity operator) is $\cal{V}=\mathcal{V}_L\otimes\mathcal{V}_R$.

Once the chiral algebra $\cal{V}_L$ is known, its representations give the states/operators in the chiral CFT. Denote the space of irreducible representations of the chiral algebra by $\mathcal{H}_a$, where $a$ is a chiral primary operator (with respect to the chiral algebra $\cal{V}_L$). These fields are also called chiral vertex operators. The defining characteristic of a RCFT is that the number of primary fields is finite.

From the topological data of the chiral primary fields of a RCFT, one can extract a unitary MTC $\mathcal{B}$, which is called the ``representation category'' of the chiral algebra, also denoted by $\mathrm{Rep}(\mathcal{V}_L)$.  The chiral CFT describes the boundary excitations of the bulk topological phase described by $\mathcal{B}$. There is a one-to-one correspondence between chiral primary fields and anyon types of the topological phase. This statement captures the essence of the bulk-boundary correspondence in these systems. Physically, the Hilbert space $\mathcal{H}_a$ is realized on the edge of a disk of the topological phase $\mathcal{B}$, where inside the disk there is a single anyonic excitation $a$. One may also imagine starting from a disk without any bulk excitation (so the boundary CFT Hilbert space is $\mathcal{H}_0$), creating a pair of anyons $a$ and $\bar{a}$ and moving $\bar{a}$ to the edge. This process changes the space of the boundary CFT to $\mathcal{H}_a$.
For this purpose, it is useful to consider the torus partition function.  We define the character associated with a chiral primary operator $a$:
\begin{equation}
	\chi_a(\tau) = \tr_{\mathcal{H}_a} e^{2\pi i\tau(L_0-\frac{c}{24})}.
	\label{}
\end{equation}
Here $\tau$ is the modular parameter of the torus, and $L_n$ is the $n$-th Virasoro generator.

The full CFT in (1+1)d contains both the chiral and anti-chiral parts. For simplicity, we assume their chiral algebras are isomorphic to each other. The complete Hilbert space is then 
\begin{equation}
	\mathcal{H}=\bigoplus_{a,b}  M_{ab}\mathcal{H}_a\otimes\ol{\mathcal{H}}_b.
	\label{eqn:CFTHib}
\end{equation}
Here $\ol{\mathcal{H}}_b$ denotes the Hilbert space of the right-moving sector, which is isomorphic to $\mathcal{H}_b$.
$M_{ab}$ are non-negative integers. For $M_{ab}>0$, the corresponding chiral and anti-chiral sectors $a$ and $b$ 
are ``paired'' to form a local primary operator, with the multiplicity given by $M_{ab}$. Thus the operator content is determined by $M_{ab}$.  The partition function then reads
\begin{equation}
	\mathcal{Z}(\tau, \bar{\tau})=\sum_a \chi_a(\tau)M_{ab} \ol{\chi}_b(\bar{\tau}).
	\label{}
\end{equation}
Moreover, the naturality theorem due to Moore and Seiberg~\cite{Moore89a} states that when the chiral algebra is maximally extended, $M$ must take the form $M_{ab}=\delta_{b, \varphi(a)}$ where $\varphi$ is an automorphism of the fusion algebra. In fact, $\varphi$ must be a braided tensor auto-equivalence of the MTC. From the bulk perspective this corresponds to the absence of condensable bosons in $\cal{B}$. 
Physical primary fields can be built out of linear combinations of products of chiral and anti-chiral primary fields (how they can be paired up to form local primaries are dictated by the partition function, due to the state-operator correspondence).  The most common choice is $M_{ab}=\delta_{ab}$, known as the diagonal CFT.

\begin{figure}[tpb]
	\centering
	\includegraphics[width=0.9\columnwidth]{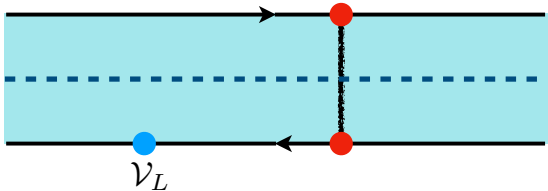}
	\caption{An illustration of the strip construction. The (blue) dot on the lower edge denotes local operators on the edge, which form the chiral algebra $\cal{V}_L$. The thick line across the strip represents a primary operator, which can be thought of as an anyon tunneling between the two edges. The dashed line denotes a gapped domain wall inside the bulk of the strip.}
	\label{fig:strip}
\end{figure}

We now discuss a physical picture of the full CFT as a (2+1)d chiral topological phase on a strip geometry~\cite{Cappelli1, Cappelli2}. This picture gives an intuitive explanation of the known classification of RCFTs~\cite{FRS1, FRS2, FRS3,FRS4}, and has also been rigorously established~\cite{KZ2017, KongCFT2019a, KongCFT2019b}. Let us consider a (2+1) topological phase, which has the topological order described by the representation category of the left chiral algebra, so the edge modes are given by the corresponding chiral CFT. Suppose that the couplings between the top and bottom edges are negligible.  We focus on the low-energy states of this strip, so that no bulk anyon excitations are allowed. In general, one should also allow a line defect (i.e. a domain wall) in the middle of the strip. This line defect may or may not be invertible. An invertible line defect in a 2D topological phase must correspond to a topological symmetry of the MTC $\mathcal{B}$. When the strip is wide and the system is viewed as two-dimensional, the only low-energy observables available are local operators on either edges, i.e. the chiral algebras. However, when the strip is viewed as a quasi-one-dimensional system, the top and bottom edges of the strip together become a non-chiral CFT~\cite{levin2013}. In this one-dimensional limit, operators which are localized in the extended direction, but straddle across the whole strip are considered local. In fact, these are the string operators that tunnel anyons between the two edges. Which types of anyons are allowed to tunnel is precisely determined by the line defect in the middle~\cite{lan2015}.
An illustration of the strip construction is shown in Fig.~\ref{fig:strip}.

It is also instructive to view the strip as a doubled topological phase $\cal{Z}(\cal{B})=\mathcal{B}\boxtimes\ol{\mathcal{B}}$, which has been unfolded along a gapped boundary condition specified by a Lagrangian algebra $\cal{A}$ in $\cal{Z}(\cal{B})$. This gapped boundary is the line defect in the strip construction.

Let us now discuss the operator content in this physical picture:
\begin{itemize}
	\item Local scaling operators on each boundary. They include spin-2 stress tensors, as well as other holomorphic/anti-holomorphic scaling operators with integer spins, such as spin-1 conserved currents of continuous symmetry. Importantly, these operators form the extended chiral algebra. 
	\item Local primary fields of the form $\Phi_{a\bar{a}}$ for a holomorphic primary $a$. They correspond to ``tunneling'' operators, which transport an anyon of type $a$ from one edge to another. If an invertible domain wall is inserted in the middle of the strip, corresponding to a topological symmetry $\varphi$ of the bulk MTC, the tunneling operator consists of creation of $a$, converted by the domain wall to $\varphi(a)$, and annihilation on the other edge. For a nontrivial $\varphi$ this strip corresponds to a non-diagonal partition function.
\end{itemize}

\section{RCFT enriched by global symmetries}
\label{sec:rcft-sym}

Now we consider RCFTs with global internal symmetry $G$. In our discussion, $G$ is the symmetry of the microscopic Hamiltonian. We focus our attention on a unitary $G$. We believe our discussions can be generalized easily to lattice translations.    

We assume that the system does not spontaneously break $G$. It is necessary to specify how local scaling operators transform under the symmetry. Since these are local, they must transform as linear representations of the symmetry group. More concretely, we distinguish three classes of local scaling operators:
 
\begin{itemize}
	\item Chiral and anti-chiral stress tensor. They by definition should transform completely trivially under any internal symmetry group $G$, i.e. $G$ commutes with the Virasoro algebra.
	\item Other chiral/anti-chiral operators in the chiral algebra, such as conserved currents of continuous symmetries. 
	\item Local primary operators of the form $\Phi_{a\bar{a}}$ for a chiral primary $a$.
\end{itemize}

It is quite common that low-energy effective field theories have more symmetries larger than the microscopic one (i.e. operators breaking the emergent symmetry are irrelevant).
Therefore we first describe the full symmetry group of the CFT. Since we mainly consider CFTs with maximally extended chiral algebra, we denote the CFT by a pair $\mathcal{V},\varphi$, and its symmetry group as $\mathrm{Sym}(\mathcal{V},\varphi)$.  Because the CFT Hilbert space is a representation space of the chiral algebra, it is useful to consider the following ``two-step'' description of global (unitary) symmetry.

First of all, symmetries act on the chiral algebra ${\mathcal{V}_L\otimes \mathcal{V}_R}$. Due to the factorization, it is sufficient to consider symmetries acting on one of them (note that this does not apply to orientation-reversing symmetries, as they must swap the two chiral algebras). In addition, as symmetries of the theory they must keep the Virasoro algebra (i.e. stress tensor) completely invariant. Let us define $\mathrm{Aut}(\cal{V})$ as the automorphism group of the chiral algebra $\cal{V}_L$.
	
It is useful to distinguish ``inner'' and ``outer'' automorphisms, denoted by $\mathrm{Inn}(\cal{V})$ and $\mathrm{Out}(\cal{V})$ respectively. Recall for finite or compact Lie groups, outer automorphisms are discrete and act nontrivially on classes of irreducible representations,  while inner automorphisms are conjugations by group elements. For chiral CFTs, analogously outer automorphisms may permute different primaries, leaving all correlation functions invariant. They can usually be identified by inspecting the CFT data given by chiral primaries. In contrast, inner automorphisms keep classes of representations invariant (but may still act nontrivially on the representation space).  
We remark that $\mathrm{Inn}(\cal{V}_L)$ is continuous if and only if the chiral algebra contains a Kac-Moody subalgebra, i.e. there are spin-1 currents, otherwise it is a finite group.

Coming back to the full non-chiral CFT, automorphisms of the full chiral algebra $\cal{V}=\cal{V}_L\otimes \cal{V}_R$ is a subgroup of $\mathrm{Aut}(\mathcal{V}_L)\times\mathrm{Aut}(\mathcal{V}_R)$ preserving modular invariance, or equivalently the Hilbert space structure in Eq.~\eqref{eqn:CFTHib}. For example, one has to restrict the outer automorphisms $\mathrm{Out}(\cal{V}_L)\times \mathrm{Out}(\cal{V}_R)$ to a diagonal subgroup.
 
Once the symmetry action on the chiral algebras is given, the action on the representation spaces (i.e. local primaries) is essentially fixed up to a phase factor. The remaining freedom will be referred to as central symmetries (since they commute with the chiral algebra). For a primary $\Phi_{a\bar{a}}$, a central symmetry acts as
\begin{equation}
	\Phi_{a\bar{a}}\rightarrow e^{i\phi_a}\Phi_{a\bar{a}}.
	\label{}
\end{equation}
Importantly, commutativity with the chiral algebra means that the phase $e^{i\phi_a}$ must satisfy
\begin{equation}
	\phi_a\phi_b=\phi_c, \text{ if }N_{ab}^c>0.
	\label{}
\end{equation}
Therefore $\phi_a$ must take the following form:
\begin{equation}
	\phi_a= M_{\cohosub{v},a},
	\label{eqn:centralsym}
\end{equation}
where $\coho{v}$ is a certain Abelian anyon in the MTC.  Using the terminology of Ref.~\cite{ChangTDL}, central symmetries are all generated by invertible Verlinde lines.
	
To summarize, the full symmetry group of a (non-chiral) CFT with maximally extended chiral algebra is an extension of $\mathrm{Aut}(\mathcal{V})$ by $\mathcal{A}$.  In other words, we have the following short exact sequence: 
\begin{equation}
	1\rightarrow \cal{A}\rightarrow \mathrm{Sym}(\cal{V},\varphi)\rightarrow \mathrm{Aut}(\cal{V})\rightarrow 1,
	\label{}
\end{equation}
where $\cal{A}$ is acted on by $\mathrm{Out}(\cal{V})$.

Now suppose the physical system that realizes the CFT has a global symmetry $G$. The action of $G$ on the CFT corresponds to a group homomorphism
\begin{equation}
	\Phi: G\rightarrow \mathrm{Sym}(\cal{V},\varphi).
	\label{}
\end{equation}
Clearly, $\Phi$ induces a homomorphism $\alpha: G\rightarrow \mathrm{Aut}(\cal{V})$, i.e. a $G$ action on the chiral algebra, which in turn induces a homomorphism $\rho: G\rightarrow \mathrm{Out}(\cal{V})$ (a $G$ action permuting local primaries). An important caveat is that $\alpha$ should be consistent with local operators transforming as \emph{linear} representations of $G$. We address this point in more detail below when the bulk interpretation is discussed.
On top of that, different choices of $\Phi$ are given by twisted homomorphisms from $G$ to $\cal{A}$, formally $[\coho{v}]\in\H^1_\rho[G, \cal{A}]$.

In the following we illustrate these general remarks with examples of WZW CFTs and the $\Z_3$ parafermion CFT.

\subsubsection{WZW CFT}
Consider a WZW CFT $\mathfrak{g}_k$ with diagonal partition function, where $\mathfrak{g}$ is a simple Lie algebra. Denote by $\cal{G}$ the corresponding simply-connected Lie group. The primaries are labeled by irreducible representations of $\cal{G}$. The corresponding MTC describes the topological order of a level $k$ Chern-Simons theory with gauge group $\mathcal{G}$.

It is well-known that the CFT has the following continuous global symmetry:
\begin{equation}
	G_\mathrm{WZW}=({\cal{G}_L\times\cal{G}_R})/{Z^\text{diag}(\cal{G})}.
	\label{eqn:gwzw}
\end{equation}
Here $Z^\text{diag}(\cal{G})$ denotes the diagonal axial center group: namely, $Z^\mathrm{diag}(\cal{G})$ consists of $(g, g^{-1})$ for each $g$ in the center of $\cal{G}$. For example, if $\mathfrak{g}=\mathfrak{su}(2)$, $G_\mathrm{WZW}=\mathrm{SO}(4)$.

Let us consider the symmetry action on the chiral algebra $\cal{V}_L$, which is the $\mathfrak{g}$ level $k$ Kac-Moody algebra, generated by the current operators $J^a$.  The inner automorphism group therefore can be identified with
\begin{equation}
	\mathrm{Inn}(\cal{V}_L)=\cal{G}/Z(\cal{G}),
	\label{}
\end{equation}
generated by the currents $J^a_0$.

In addition, discrete symmetries which permute primary fields are given by the group of outer automorphisms of the Lie group, which coincides with the symmetry group of the Dynkin diagram.   For instance, for $\mathfrak{g}=\mathfrak{so}(n)$ one finds that $\mathrm{Out}(\mathfrak{so}(n)_k)=\Z_2$ (except for $n=8$, $\mathfrak{so}(8)$ has an $S_3$ outer automorphism group), thus together with $\mathrm{Inn}(\cal{V}_L)=\mathrm{SO}(n)$ we have
\begin{equation}
	\mathrm{Aut}(\mathfrak{so}(n)_k)=\mathrm{O}(n).
	\label{}
\end{equation}
Or for $\mathfrak{g}=\mathfrak{su}(n)$, we find $\mathrm{Inn}(\mathfrak{su}(n)_k )=\mathrm{PSU}(n)$ and for $n>2$ there is a $\Z_2$ charge conjugation symmetry, so $\Aut(\mathfrak{su}(n)_k)=\mathrm{PSU}(n)\rtimes \Z_2$.  Notice that what we have written down so far are automorphisms of the left (or right) chiral algebra. If both left and right chiral algebras are considered, then
\begin{equation}
	\big[\cal{G}_L/Z(\cal{G}_L)\times\cal{G}_R/Z(\cal{G}_R) \big]\rtimes \mathrm{Out}(\cal{G}),
	\label{}
\end{equation}
where now elements of $\mathrm{Out}(\cal{G})$ act diagonally on both chiral algebras.

Notice that this differs from the actual symmetry group, whose connected component is given in Eq.~\eqref{eqn:gwzw}. The difference is exactly the center $Z(\mathcal{G}_L)$ (or $Z(\mathcal{G}_R)$). In other words, if the symmetry action on the chiral algebra is known, the symmetry transformations on all the primaries are known up to the center group. The remaining degree of freedom is a homomorphism ${v}:G\rightarrow Z(\cal{G})$. That is,  on a primary labeled by an irrep $R$, one attaches a one-dimensional representation defined by 
\begin{equation}
	\phi_R(\mathbf{g})=R\big({v}(\mathbf{g})\big).
	\label{eqn:wzw-center-sym}
\end{equation}
As we have shown in Eq.~\eqref{eqn:centralsym}, such a one-dimensional rep. must be generated by braiding with Abelian anyons in the corresponding MTC. Indeed, the group $\mathcal{A}$ of abelian anyons is naturally isomorphic to the centre $Z(\mathcal{G})$, with one notable exception being $(E_8)_2$.

%If the symmetry group $G$ acts chirally, e.g. $G$ is identified with $\cal{G}_L/Z(\cal{G})$, then there is a chiral anomaly. However, when $Z(\cal{G})$ is nontrivial, if one naively takes all the field content of the original WZW CFT, certain local primaries do not form faithful representations of $G$. For instance, if $\mathfrak{g}=\mathfrak{su}(2)$ and take $G=\mathrm{SO}(3)_L$, then all primaries with half-odd integer spin are projective representations of SO(3)$_L$. In other words, one can only keep the integer spin primaries. However, this does not always form a consistent CFT. ``SO(3)$_k$'' CFTs only exist for $k\equiv 0\,(\mathrm{mod }4)$, and can be obtained from the $\mathfrak{su}(2)_k$ WZW CFT by extending the chiral algebra with the spin $k/2$ primary. 

\subsubsection{$\Z_3$ parafermion CFT}
The $\Z_3$ parafermion (PF) CFT can be realized as the coset $\mathrm{SU}(2)_3/\mathrm{U}(1)$, or from the minimal model $\mathcal{M}(6,5)$ by extending the Virasoro algebra by a spin-$3$ chiral primary. The resulting chiral algebra is known as the $W_3$-algebra~\cite{ZamolodchikovW}. We denote the spin-$3$ chiral primary by $W(z)$. The OPE of $W$ is quite complicated, with the leading terms being $W(z)W(w)\sim \frac{2/3}{(z-w)^6}+\frac{2T(w)}{(z-w)^4}+\cdots$ (a few more singular terms are omitted). The only nontrivial discrete unitary symmetry on the chiral algebra is given by
\begin{equation}
	W\rightarrow -W.
	\label{}
\end{equation}
This is nothing but the charge conjugation symmetry that sends a parafermionic field $\psi(z)$ to $\psi^\dag(z)$. 

The $\Z_3$ central symmetry of the CFT is generated by parafermionic fields $\psi$. Together with the $\Z_2$ symmetry they form the $\mathbb{S}_3$ group.

\subsection{$(2+1)d$ bulk interpretation}
It is natural to interpret the symmetry action on a CFT from the perspective of the $(2+1)$d bulk. Recall that global $G$ action in a (1+1)d CFT can be specified by two maps: $\alpha: G\rightarrow \mathrm{Aut}(\cal{V})$ and $[\coho{v}]\in \cal{H}^1_\rho[G, \cal{A}]$ where $\rho: G\rightarrow \mathrm{Aut}\big(\mathrm{Rep}(\cal{V})\big)$ is induced by $\alpha$.

Let us first consider the symmetry action on the chiral algebra, given by $\alpha: G\rightarrow \mathrm{Aut}(\cal{V})$. 
Due to the holomorphic factorization of the chiral algebras, $\alpha$ can be restricted to the left or the right algebra. Suppose we have a group homomorphism $\alpha_L:G\rightarrow \mathrm{Aut}(\cal{V}_L)$. Via the bulk-boundary correspondence, the homomorphism $\alpha_L$ encodes the symmetry action on local operators on the chiral edge of the bulk topological phase $\mathrm{Rep}(\cal{V}_L)$.     We argue that $\alpha_L$ in fact completely determines the symmetry action on anyons, i.e. the symmetry-enriched topological order in $\cal{C}=\mathrm{Rep}(\cal{V}_L)$. Namely, $\rho, U$ and $\eta$ are completely determined.

Let us elaborate on this statement.  Denote $\cal{C}=\mathrm{Rep}(\cal{V}_L)$, whose anyon types are labeled as $a,b,c,\dots$ in one-to-one correspondence with chiral primaries of the chiral CFT. $\alpha_L$ induces a group homomorphism from $G$ to $\mathrm{Out}(\cal{V}_L)$. Each element of $\mathrm{Out}(\cal{V}_L)$ corresponds to an element in $\mathrm{Aut}(\cal{C})$, although not the other way around. In other words, there is a canonical embedding $\mathrm{Out}(\cal{V}_L)\subset \mathrm{Aut}(\cal{C})$. Therefore $\alpha_L$ defines uniquely a homomorphism $\rho: G\rightarrow \mathrm{Aut}(\cal{C})$.

It is less obvious that that $\alpha_L$ also determines $U_\mb{g}(a,b;c)$ and $\eta_a(\mb{g,h})$.   Here we present physical arguments for why this is the case. In short, this is because $\alpha_L$ defines the symmetry action on local operators in the chiral edge theory. 

Each chiral primary/anyon type $a$ is associated with a Hilbert space $\cal{H}_a$ in the chiral CFT, which is the edge Hilbert space on a disk with an anyon $a$ inside. As $\{\cal{H}_a\}$ are the representation spaces of the chiral algebra $\cal{V}_L$, any automorphism of $\cal{V}_L$ induces a unitary map on $\{\cal{H}_a\}$. More specifically, an inner automorphism induces a unitary transformation  on each $\cal{H}_a$. An outer automorphism maps $a$ to $a'$, thus inducing a map between isomorphic spaces $\cal{H}_a$ and $\cal{H}_{a'}$. These unitary transformations provide a concrete realization of the symmetry localization in Eq.~\eqref{eqn:Rdecomp}, and one can then directly compute the $\eta$ symbols following the definition. 

It is also instructive to define $\eta$ in the operator language. An anyon must be created by a non-local string operator, where the string itself commutes with Hamiltonian at low energy. In other words, it corresponds to a topological defect line in the chiral CFT. In the Hamiltonian formalism, let us denote the chiral primary operator by $V_a(x)$, which is a semi-infinite string operator. Under an outer automorphism that sends $a$ to $a'$, $V_a(x)$ should be transformed to $V_{a'}(x)$, up to a local operator at $x$. Via the state-operator map, this local unitary operator corresponds to the map between $\cal{H}_a$ and $\cal{H}_{a'}$.

Next we consider $U$, the symmetry action on splitting spaces. Intuitively, the splitting space $V^{ab}_c$ is the equivalence classes of local operators that split an anyon $c$ to two anyons $a$ and $b$. Since $\alpha_L$ defines the action on local operators, the symmetry action can be computed once the splitting operator is known.

In Appendix~\ref{app:free-boson} we present more precise definitions of $U$ and $\eta$ in terms of CFT operators, and explicitly compute them for the charge-conjugation symmetry in a $\U_N$ CFT (i.e. compactified chiral free boson).

We remark that while we define $U$ and $\eta$ for a chiral CFT, essentially the same can be done for a full non-chiral theory, where now chiral primaries become topological defect operators.

Here we illustrate the definition of $\eta$ using the example of WZW CFTs. Neglecting the outer automorphisms for the moment, the symmetry action on the chiral algebra is given by the group homomorphism $\alpha: G\rightarrow \cal{G}/Z(\cal{G})$. Consider states in $\mathcal{H}_R$ where $R$ is an irreducible representation of $\mathcal{G}$. For $\mb{g}\in G$, the group homomorphism fixes a unitary transformation $R(\mb{g})\equiv R\big(\alpha(\mb{g})\big)$ on $\mathcal{H}_R$ (as the representation $R$ of $\cal{G}$) up to the center $Z(\cal{G})$, which is just the ambiguity of phase factors given in Eq.~\eqref{eqn:wzw-center-sym}. In other words, the chiral primary $R$ may transform projectively under $G$:
\begin{equation}
	R(\mb{g})R(\mb{h})=\phi_R(w(\mb{g,h}))R(\mb{gh}),
	\label{}
\end{equation}
where $w(\mb{g,h})\in Z(\cal{G})$. Through the bulk-boundary correspondence, $\phi_R(w(\mb{g,h}))$ is also the projective phase on the bulk anyon labeled by $R$. As a result, the fractionalization class $[\coho{w}]\in \H^2[G,\cal{A}]$ is determined through the canonical map between $Z(\cal{G})$ and $\mathcal{A}$.

A useful corollary is that if $\alpha_L=\mathds{1}$, i.e. the symmetry group commutes with the entire chiral algebra, then one may set $\rho\equiv\mathds{1}$, and $U,\eta$ all to $1$. The symmetry thus acts completely trivially on anyons. 

We remark that most familiar examples of chiral symmetry-enriched RCFTs with nontrivial symmetry fractionalization class $\eta$ are WZW CFTs. More exotic examples without Kac-Moody algebra do exist, which are discussed in Sec.~\ref{sec:proj_no_KM}.  

Now let us come back to the full (1+1)d CFT. As we have mentioned, an additional piece of information needed is a Lagrangian algebra $\cal{L}$ in the doubled MTC $\cal{Z}\big(\mathrm{Rep}(\cal{V}_L)\big)$. It is necessary that $\alpha_L$, and thus the SET order in $\mathrm{Rep}(\cal{V}_L)$, allows $\cal{L}$ to be symmetric. When the corresponding $\rho$ is trivial, it means that condensed anyons in $\cal{L}$ do not carry any projective representations. For example, for $a$ invariant under $G$, $\Phi_{a\bar{a}}$ transform as linear representations of $G$, as expected. Schematically let us write
\begin{equation}
	\mb{g}\Phi_{a\bar{a}}\mb{g}^{-1}=R_a(\mb{g})\Phi_{\rho(a)\ol{\rho(a)}},
	\label{}
\end{equation}
where $R_a(\mb{g})$ is a unitary transformation. Additional indices are suppressed. $R_a(\mb{g})$ can then be modified with phase factors: $R_a(\mb{g})\rightarrow M_{a,\cohosub{v}(\mb{g})}R_a(\mb{g})$. In other words, $\cal{H}^1_\rho[G,\cal{A}]$ provides a torsor over distinct symmetry-enriched CFTs with the same $\alpha_L$. Clearly, this should be identified with the different charged condensations classified in Sec.~\ref{sec:SETboundary}.

\subsection{Absolute and relative 't Hooft anomalies}
First we consider 't Hooft anomalies generally for global symmetries in a RCFT. 

On general grounds, we expect that the anomaly $3$-cocycle can be derived once the action of $G$ on local scaling operators in the CFT is known. We explain how such a computation can be done in principle in Sec.~\ref{sec:general-anomaly}. However, while the procedure is well-defined, the relation between the prescription and local data is rather obscure. 

In the following, we instead study the following question: given a RCFT and a symmetry group $G$, consider the possible 't Hooft anomalies that different symmetry-enriched RCFTs may exhibit. We have shown that the full symmetry group of the RCFT is an extension of $\mathrm{Aut}(\cal{V})$ by $\cal{A}$, and thus a $G$ action on local operators consists of two homomorphisms $\alpha: G\rightarrow \mathrm{Aut}(\cal{V})$ and $[\coho{v}]\in \H^1_\rho[G, \cal{A}]$.  In principle, the anomaly 3-cocycle is a function of $\alpha$ and $[\coho{v}]$, and we define the relative anomaly as $\omega(\alpha, [\coho{v}_0\cdot\coho{v}])\cdot\omega(\alpha, [\coho{v}_0])^{-1}$.
We provide an explicit formula to compute the ``relative'' 't Hooft anomaly that only uses algebraic data.

\subsubsection{'t Hooft anomaly from topological defect lines}
\label{sec:general-anomaly}
We first review the general method to compute 't Hooft anomaly in a symmetry-enriched CFT in the language of Euclidean quantum field theory, following Ref.~\cite{ChangTDL}. Related discussions can be found in Refs.~\cite{FrohlichNPB2006,Aasen2016,BultinckPRL2018}.

In a quantum field theory, global symmetries are implemented by invertible topological operators of codimension 1 (which may be viewed as the spacetime trajectory of a symmetry defect). In (1+1)d, they are invertible topological defect lines (TDL)~\cite{ChangTDL} supported on an oriented path that commute with the stress tensor. A TDL can end on a point-like defect operator. By the state-operator correspondence, for each $\mb{g}\in G$ there is a Hilbert space $\mathcal{H}_\mb{g}$, namely the theory on $S^1$ with the boundary condition twisted by $\mb{g}$. 

\begin{figure}[tpb]
	\centering
	\includegraphics[width=\columnwidth]{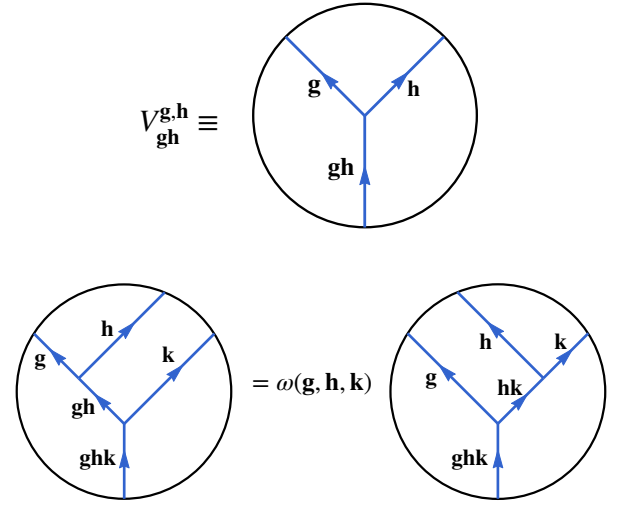}
	\caption{Top panel: illustration of a three-way junction of TDLs which define the space $V^{\mb{g,h}}_\mb{gh}$. Bottom: definition of the 3-cocycle $\omega(\mb{g,h,k})$. }
	\label{fig:tdl}
\end{figure}
TDLs may form junctions and more general networks. Consider the simplest three-way junction of three TDLs labeled by $\mb{g,h}$ and $\ol{\mb{gh} }$. On a disk, such a configuration defines a Hilbert space $\cal{H}^{\mb{g,h}}_\mb{gh}$, isomorphic to the untwisted Hilbert space in the identity sector. Let us label the ground state subspace of the Hilbert space as $V^{\mb{g,h}}_\mb{gh}$~\footnote{Since TDLs commute with the stress tensor, the Hilbert space $\cal{H}^{\mb{g,h}}_\mb{gh}$ still forms a representation of the left and right Virasoro algebras. Following Ref.~\cite{ChangTDL} we define the ground state space to be the space of weight-$(0,0)$ states. This choice of the weight-$(0,0)$ state also fixes the junction operator itself through state-operator correspondence. Such a junction is defined as a topological junction in Ref.~\cite{ChangTDL}.}. The definition is illustrated in Fig.~\ref{fig:tdl}. The discussion can be generalized to any $k$-way junctions straightforwardly. 

Now consider a four-way junction ${\mb{g,h,k}}$ and $\ol{\mb{ghk}}$. For the same reason, the junction defines a one-dimensional space $V^{\mb{g,h,k}}_\mb{ghk}$. There are two ways to draw the TDLs inside the disk, which should lead to the same state space. Thus they can only differ by a phase factor $\omega(\mb{g,h,k})$, which is the anomaly $3$-cocycle~\cite{ChangTDL}. See Fig. \ref{fig:tdl} for an illustration.

The procedure to extract $\omega$ outlined above may be viewed as a special case of the Else-Nayak prescription of extracting $3$-cocycles for a set of unitary operators that represent the symmetry group~\cite{ElsePRB2014} (see also Ref. \cite{KawagoeSPT}). Let us briefly explain the relation. In the Else-Nayak prescription, global symmetries are represented by (locality-preserving) unitary operators  $\{U(\mb{g})\}_{\mb{g}\in G}$ acting on a 1D lattice model. One first considers the restriction of a global symmetry unitary $U(\mb{g})$ to an open region $M$ of the lattice, denoted by $U_M(\mb{g})$~\footnote{The restriction $U_M$ is defined such that for any local operator $O$ supported inside $M$, we have $UOU^{-1}=U_M O U_M^{-1}$}. For convenience, we take $M$ to be the half line with one end point $e$. If the low-energy theory is a CFT, the scaling limit of the restricted unitary defines a symmetry defect operator in the CFT (this is not unique, but does not affect final result). The restricted operators in general obey the group multiplication law only up to a local operator at $e$: 
\begin{equation}
	U_M(\mb{g})U_M(\mb{h})=\Omega_e(\mb{g,h})U_M(\mb{gh}).
	\label{}
\end{equation}
For CFT the local operator $\Omega_e(\mb{g,h})$ corresponds to a state in $\cal{H}^{\mb{g,h}}_\mb{gh}$ under the state-operator map~\footnote{In the CFT, a canonical choice for $\Omega_e$ is made by demanding that the junction is topological. This is convenient but not essential}. The associativity condition of $\Omega_e$ is satisfied up to a phase, which is the 3-cocycle. This is essentially the operator version of the CFT definition as illustrated in Fig. \ref{fig:tdl}.

\subsubsection{Relative 't Hooft anomaly}
We define the relative 't Hooft anomaly to be the \emph{difference} between the 't Hooft anomalies of two symmetry-enriched RCFTs, whose symmetry actions differ by $[\coho{v}]\in \H^1_\rho[G, \cal{A}]$ (with the same $\alpha$). In the language of TDL, it means that each defect line is dressed by an additional invertible Verlinde line. 

We now derive a formula for the relative anomaly using the correspondence between symmetry-enriched (1+1)d CFT and (2+1)d doubled SET phase. First we assume that the theory has a diagonal partition function, so the CFT can be labeled by the chiral algebra $\cal{V}_L$. Denote $\cal{C}=\mathrm{Rep}(\cal{V}_L)$. As described in the previous section, the group homomorphism $\alpha_L$ uniquely defines a (2+1)d SET phase, which we denote by $\cal{C}_G^\times$. Similarly, $\alpha_R$ defines an SET phase with the conjugate topological order, denoted by $\ol{\cal{C}}_G^\times$. Together they form a double-layer SET phase. The two must have opposite symmetry actions on anyons, in order to have a diagonal, symmetry-preserving Lagrangian algebra Eq.~\ref{eqn:L-diag}. Condensing the Lagrangian algebra results in an SPT phase. Or equivalently, the bilayer SET phase should admit a fully gapped, symmetric boundary to an SPT phase. In order to derive the relative anomaly, without loss of generality we can assume that the two SET phases are exactly conjugate to each other. The SPT phase is given in Eq.~\ref{eqn:anomaly2}, which is the formula for the relative anomaly. We reproduce the formula here:
\begin{equation}
	\begin{split}
	\omega(\mb{g,h,k})=& F^{\cohosub{v}(\mb{g}), {}^\mb{g}\cohosub{v}(\mb{h}), {}^\mb{gh}\cohosub{v}(\mb{k})}\\
	&\cdot U_\mb{g}^{-1}\big({}^\mb{g}\coho{v}(\mb{h}), {}^\mb{gh}\coho{v}(\mb{k})\big) \eta_{ {}^\mb{gh}\cohosub{v}(\mb{k})}(\mb{g,h}).
	\end{split}
	\label{eqn:anomaly2}
\end{equation}

While in general our formula only computes the relative anomaly, in some cases one can identify non-anomalous CFTs as a reference to get absolute anomalies.

The first case is when $G$ commutes with the whole chiral algebra, i.e. $\alpha=\mathds{1}$. As we have argued, the bulk SET phase has trivial symmetry action, and when all primaries transform trivially under $G$ the theory is obviously non-anomalous ($G$ essentially does not act at all).

Alternatively, $G$ may act nontrivially on the chiral algebra, but the transformations of left and right chiral algebras are conjugate to each other. In terms of $\alpha$, it implies that $\alpha_L$ and $\alpha_R$ are basically identical (with respect to the isomorphism between $\cal{V}_L$ and $\cal{V}_R$).  In this case, the CFT can be constructed by putting the corresponding chiral bulk SET phase on a strip, or equivalently as the boundary of the doubled SET phase.

In these cases, there is a ``canonical'' symmetry-enriched CFT which is non-anomalous. Thus the relative anomaly computed from the reference theory becomes the absolute anomaly.

We note that Ref. \cite{relativeanomalySET} computed relative 't Hooft anomaly in (2+1)d SET phases. There tri-junctions of symmetry defect surfaces can be decorated by Abelian anyon line operators to change the symmetry fractionalization class and possibly the 't Hooft anomaly. As mentioned in Sec. \ref{sec:SETboundary}, our results are closely related to the relabeling of symmetry defects by fusing Abelian anyons in the (2+1)d theory. 

\section{Examples}
In this section we study various examples. We first consider global symmetries of $\mathrm{SU}(2)_k$ CFTs and the closely related unitary minimal models, including all non-diagonal theories. One motivation is to illustrate how our approach to symmetry-enriched CFTs works when the partition function is non-diagonal (and not necessarily maximally extended). We show that minimal models (including the non-diagonal ones) are always non-anomalous, thus CFTs with nontrivial 't Hooft anomalies must have $c\geq 1$. We derive general constraints on 't Hooft anomaly when the global symmetry commutes with the whole chiral algebra. We then classify symmetries in diagonal WZW CFTs. Finally, we study Lieb-Schulz-Mattis anomaly in translation-invariant spin chains.

\subsection{Global symmetries of SU(2)$_k$}
In this section we analyze global symmetries of SU(2)$_k$ CFTs, in particular those with non-diagonal partition functions.
In the following we denote chiral primaries of SU(2)$_k$ theory by $V_j$, where $j$ is the SU(2) spin $j=0,\frac{1}{2},\dots, \frac{k}{2}$. The corresponding characters are denoted by $\chi_j(\tau)$. 

Modular-invariant partition functions of SU(2)$_k$ CFTs are completely classified, known as the ADE classification~\cite{Cappelli1987}. Besides the diagonal partition functions, there are six classes of non-diagonal theories. In all cases, the chiral algebra contains an $\mathfrak{su}(2)$ Kac-Moody (sub)algebra, so there is at least $\mathrm{SO}(3)_L\times\mathrm{SO}(3)_R$ symmetry.
 In a few cases the Kac-Moody algebra is further enlarged ($D_4, E_6$ and $E_8$ modular invariants). We will pay attention to additional discrete global symmetries that are not part of the continuous symmetry group.

\begin{enumerate}
	\item $A_n$ series for all $k$: the global symmetry is $\mathrm{SU}(2)_L\times\mathrm{SU}(2)_R/\Z_2=\mathrm{SO}(4)$. In particular, the $\Z_2$ center symmetry generated by the Verlinde line $V_{\frac{k}{2}}$ is anomalous/non-anomalous for odd/even $k$. 
	\item $D_{2n+2}$ series for $k=4n$: This series arise from the extension of the chiral algebra by the $V_{\frac{k}{2}}$ primary, which has conformal weight $\frac{k}{4}$ . The partition function reads
		\begin{equation}
			\sum_{\substack{m=0\\ m\in\Z}}^{n-1}|\chi_m+\chi_{2n-m}|^2+2|\chi_{n}|^2.
			\label{}
		\end{equation}
		Since the $\Z_2$ center symmetry before extension is generated by the $V_{\frac{k}{2}}$ Verlinde line, the symmetry is gone now as only those that are neutral under the $V_{\frac{k}{2}}$ line are left. Thus the continuous SO(4) symmetry is reduced to $\mathrm{SO}(3)_L\times \mathrm{SO}(3)_R$.
		Interestingly, the new CFT has an additional $\Z_2$ symmetry swapping the two primary operators that split from the $V_{\frac{k}{4}}$ primary in SU(2)$_k$. We remark that this symmetry does not commute with the extended chiral algebra (acting as $V_{\frac{k}{2}}\rightarrow -V_{\frac{k}{2}}$). It is by construction non-anomalous, since orbifolding the symmetry gives back SU(2)$_k$. 
	\item $D_{2n+1}$ series for $k=4n-2$, with $n\geq 2$: This series arises because of a topological symmetry $V_j\leftrightarrow V_{\frac{k}{2}-j}$ (notice that it is not an actual symmetry of the CFT). The partition function reads
		\begin{equation}
			\sum_{\substack{m=0\\ m\in\Z}}^{2n-1}|\chi_m|^2 + \sum_{\substack{0<j<n-1\\ j\in\Z+\frac{1}{2}}}(\chi_{j}\chi_{2n-1-j}^*+\text{c.c}) +|\chi_{n-\frac{1}{2}}|^2.
			\label{}
		\end{equation}
		The symmetry group is the same SO(4) as the diagonal theories.

	\item $E_6$ for $k=10$: The chiral algebra is extended by the chiral primary $V_3$. The partition function reads
	\begin{equation}
|\chi_0+\chi_3|^2 + |\chi_{\frac{3}{2}}+\chi_{\frac{7}{2}}|^2+|\chi_2+\chi_5|^2.
			\label{}
		\end{equation}
		Since $V_3$ has conformal weight $1$, we expect that the Kac-Moody algebra is enlarged as additional spin-$1$ currents are added. Indeed, this partition function is actually the same as that of Spin(5)$_1$.
		%{the $j=3$ chiral primary has $h=1$, thus adding 7 spin-$1$ currents to the chiral algebra.  The total number of spin-$1$ fields is $3+7=10$). 
	\item $E_7$ for $k=16$: the partition function is
	  \begin{equation}
		  \begin{split}
		  |\chi_0&+\chi_8|^2 + |\chi_2+\chi_6|^2 +  |\chi_3+\chi_5|^2+|\chi_4|^2\\
		  &+\chi_4(\chi_1^*+\chi_7^*)+\chi_4^*(\chi_1+\chi_7).
		  \end{split}
		  \label{eqn:su2E7}
	  \end{equation}
	  The modular invariant can be understood in two steps. First, since $k$ is a multiple of $4$ one can construct a $D_{10}$ invariant, by adding $V_8$ to the chiral algebra.  The resulting SO(3)$_{16}$ CFT has 6 primaries, which can be labeled using the corresponding SU(2) spin in the parent theory: $0,1,2,3, 4_{\pm}$. The partition function at this point is
		\begin{equation}
			|\chi_0+\chi_8|^2+|\chi_1+\chi_7|^2+|\chi_2+\chi_6|^2+|\chi_3+\chi_5|^2 + 2|\chi_4|^2.
			\label{}
		\end{equation}
		As we have discussed previously, SO(3)$_{16}$ CFT has an additional $\Z_2$ symmetry that swaps $4_+$ with $4_-$. 

		Second, notice the corresponding MTC has a new topological symmetry, swapping $4_+$ (or $4_-$) with $1$, which can be used to construct the modular invariant in Eq.~\eqref{eqn:su2E7}. However, the construction of the modular invariant apparently breaks the symmetry between $4_+$ and $4_-$.  Thus the global symmetry is just $\mathrm{SO}(3)_L\times \mathrm{SO}(3)_R$.
	\item $E_8$ for $k=28$: The chiral algebra is extended by multiple chiral primaries. The partition function is
		\begin{equation}
		|\chi_0+\chi_5+\chi_9+\chi_{14}|^2+|\chi_3+\chi_6+\chi_8+\chi_{11}|^2.
			\label{}
		\end{equation}
		For the MTC, there is a condensable algebra $0+5+9+14$.  To understand this case, we perform the condensation in two steps: first condense $14$ to obtain SO(3)$_{28}$, where a new $\Z_2$ symmetry that swaps $7_\pm$ emerges. However, in the second step $5$ (which is identified with $9$) condenses, confining both $7_\pm$ and the new $\Z_2$ symmetry is no longer present.  In fact, the final theory is $(G_2)_1$ because the $V_5$ has spin 1.
\end{enumerate}

The results derived in this section are summarized in Table.~\ref{tab:su2-sym}, and are in agreement with Ref.~\cite{LienartNPB2001}.

\begin{table}
	\centering
	\begin{tabular}{|c|c|c|c|}
		\hline
		SU(2) level & ADE label & $\mathrm{Aut}(\mathcal{V})$ & Center\\
		\hline
		$k>0$   & $A_{k+1}$ & $\mathrm{SO}(3)_{L/R}$ & $\Z_2$\\
		$k=4n, n>1$ & $D_{2n+2}$ & $\mathrm{SO}(3)_{L/R}\times\Z_2$ & \\
		$k=4$ & $D_4$ & $\mathrm{PSU}(3)_{L/R}\rtimes\Z_2$ & $\Z_3$\\
		$k=4n-2, n>1$ & $D_{2n+1}$ & $\mathrm{SO}(3)_{L/R}$ & $\Z_2$\\
		$k=10$ & $E_6$ & $\mathrm{SO}(5)_{L/R}$ & $\Z_2$\\
		$k=16$ & $E_7$ & $\mathrm{SO}(3)_{L/R}$ & \\
		$k=28$ & $E_8$ & $(G_2)_{L/R}$ & \\
		\hline
	\end{tabular}
	\caption{Global symmetries of SU(2)$_k$ CFTs. $G_{L/R}$ is short for $G_L\times G_R$.}
	\label{tab:su2-sym}
\end{table}

\subsection{Symmetries in minimal models}
\label{sec:mm}

Now we study symmetries in minimal models, which are the only unitary CFTs with $c<1$. We label them as $\cal{M}(m+1, m)$ where $m\geq 3$, with central charge $c=1-\frac{6}{m(m+1)}$. Primary operators are labeled by two integers $(r,s)$ where $1\leq r\leq m-1, 1\leq s\leq m$, with the identification $(r,s)\equiv (m-r, m+1-s)$, and the conformal weight are
\begin{equation}
	h_{r,s}=\frac{[(m+1)r-ms]^2-1}{4m(m+1)}.
	\label{}
\end{equation}
More details about minimal models can be found in, e.g. Ref. \cite{difrancesco}.

First we consider diagonal theories. Since the chiral algebra of the minimal models is just the Virasoro algebra, any unitary symmetry must commute with the chiral algebra, i.e. they are central. As discussed in Sec.~\ref{sec:rcft-sym},  central symmetries correspond to simple currents in the chiral CFT. There is only one simple current in the minimal model $\cal{M}(m+1,m)$, namely the $(m-1,1)$ which necessarily has $\Z_2$ fusion rule. Thus the only nontrivial symmetry acts on the primary $(r,s)$ by a sign $(-1)^{(m+1)r+ms+1}$. This is of course expected from the Ginzburg-Landau description as a $\Z_2$ multi-critical point.

Now let us turn to non-diagonal theories, which have a well-known ADE classification~\cite{Cappelli1987, Kato1987}. They can be understood most easily in terms of the coset construction:
\begin{equation}
	\cal{M}(k+3,k+2)=\frac{\mathrm{SU}(2)_{k}\times\mathrm{SU}(2)_1}{\mathrm{SU}(2)_{k+1}}.
	\label{eqn:coset-mm}
\end{equation}
In fact, non-diagonal partition functions of a minimal model are labeled by a pair of ADE labels for the corresponding SU(2) factors in Eq.~\eqref{eqn:coset-mm}, one of which must be from the $A$ series. Then the global symmetry of the minimal model can be read off from that of the non-diagonal  SU(2) modular invariant, once the continuous part of the automorphism group is modded out.

In the following we work out the symmetries in the $(A_{m-1}, D_{\frac{m+3}{2}})$ theory for $m\equiv 1\,(\text{mod }4)$.
Since $h_{m-1,1}$ is an integer the chiral primary $(m-1,1)$ can be used to extend the Virasoro algebra, resulting in the $(A_{m-1}, D_{\frac{m+3}{2}})$ partition functions. A relevant fusion rule is $(m-1,1)\times (r,s)=(m-r,s)\equiv (r, m+1-s)$. So under fusion, there is exactly fixed point $(1, \frac{m+1}{2})$ for odd $m$ and $(\frac{m}{2},1)$ for even $m$.  
They ``split'' into two primaries of the same dimension with respect to the extended chiral algebra. The resulting theory has a new $\Z_2$ symmetry that swaps these two primaries. 

For $m=5$, the extension leads to additional simple currents enlarging the symmetry. This is because $(1,3)$ has quantum dimension $2$ and $h=2/3$.  It splits into two simple currents, which form a $\Z_3$ group.  Together with the $\Z_2$ symmetry that swaps the two simple currents,  they form an $\mathbb{S}_3$ symmetry group. The resulting theory is the $\Z_3$ parafermion CFT, where the simple currents are the parafermions.   Physically it is realized by the critical  3-state Potts models.
	%For $m=6$, $(3,1)$ has quantum dimension $2$ and $h=4/3$.  It splits into two simple currents, which form a $\Z_3$ group. realized by tricritical 3-state Potts model.
 
We note that Ref.~\cite{Ruelle1998} classified all non-anomalous symmetries based on modular invariance. Here we have essentially reproduced their results without assuming that symmetries are non-anomalous.

\subsection{$G$ commuting with the chiral algebra}
\label{sec:Gcentral}
Now we consider more general CFTs where $G$ commutes with the chiral algebra. In this case, the 3-cocycle is given by:
\begin{equation}
	\omega(\mb{g,h,k})=F^{\cohosub{v}(\mb{g}), \cohosub{v}(\mb{h}), \cohosub{v}(\mb{k})}.
	\label{eqn:omega-just-F}
\end{equation}
It turns out that a few quite general statements can be made about 3-cocycles of this form.

First of all, it is known that there always exists a gauge in which the F symbols of Abelian anyons in a MTC take values $\pm 1$~\cite{Moore89b}. Therefore, the resulting 3-cocycle must be an order-2 element in $\H^3[G, \U]$. Without loss of generality, we may assume that the symmetry group is $\Z_2^k$. Using the K\"unneth formula one can easily find~\cite{Propitius1995}
\begin{equation}
	\H^3[\Z_2^k, \U]=\Z_2^{k+\binom{k}{2}+\binom{k}{3}}.
	\label{eqn:h3z2}
\end{equation}
In fact, it is sufficient to consider $k\leq 3$, as for $k>3$ the 3-cocycles can be built from those of subgroups generated by fewer than four $\Z_2$'s, which explains the combinatorial factors in Eq.~\eqref{eqn:h3z2}.

For $k=1$, it is known that the 't Hooft anomaly can be realized by center symmetries of many WZW models~\cite{AnomalyBound}.  We now show that for $k>1$, in fact even for $k=2$, not every cohomology class in Eq.~\eqref{eqn:h3z2} can be realized by symmetries commuting with the entire chiral algebra.

First consider $k=2$. Denote $\Z_2^2=\{1, \mb{g,h,gh}\}$, then the following three invariants completely characterize the cohomology classes: $\omega(\mb{g,g,g}), \omega(\mb{h,h,h}), \omega(\mb{gh,gh,gh})$. They can take $\pm 1$ independently, which give all $2^3=8$ classes. However, given that $\coho{v}(\mb{g}), \coho{v}(\mb{h})$ are self-dual Abelian anyons, and $F^{a,a,a}=\theta_a^2$ for self-dual $a$, we find that
\begin{equation}
	F^{\cohosub{v}(\mb{gh}), \cohosub{v}(\mb{gh}), \cohosub{v}(\mb{gh})}=\theta_{\cohosub{v}(\mb{gh})}^2=
	\theta_{\cohosub{v}(\mb{g})}^2\theta_{\cohosub{v}(\mb{h})}^2M_{\cohosub{v}(\mb{g}), \cohosub{v}(\mb{h})}^2.
	\label{}
\end{equation}
Notice that $M_{\cohosub{v}(\mb{g}), \cohosub{v}(\mb{h})}^2=1$, we obtain the following constraint:
\begin{equation}
	\omega(\mb{g,g,g})\omega(\mb{h,h,h})\omega(\mb{gh,gh,gh})=1.
	\label{}
\end{equation}
Therefore, out of the eight classes in $\H^3[\Z_2^2, \U]$, only four can be realized.

Now we go to $k=3$. One can build 3-cocycles in subgroups of $\Z_2^3$, but there are so-called ``type-III'' cocycles that can only be defined for the whole $\Z_2^3$ group~\cite{Propitius1995}. Symmetries with such 3-cocycles are relevant for the phase transition between the nontrivial $\Z_2\times Z_2$ SPT and the trivial phase in (1+1)d~\cite{Bridgeman2017}. 
 The defining feature of the type-III cocycle is the following: Denote the generators of the three $\Z_2$ subgroups by $\mb{g,h,k}$.
 With respect to the first $\Z_2$ generator $\mb{g}$, the slant product of the cocycle $i_\mb{g}\omega$ is a nontrivial 2-cocycle of the remaining $\Z_2\times\Z_2$ group, detected by the following indicator:
\begin{equation}
	\frac{i_\mb{g}\omega(\mb{h,k})}{i_\mb{g}\omega(\mb{k,h})}=-1.
	\label{eqn:type3-inv}
\end{equation}

We now show that 3-cocycles of the form Eq.~\eqref{eqn:omega-just-F} can not be type-III.
This essentially follows from the hexagon equations: let $a,b,c$ be three Abelian anyons, then the hexagon equation gives
\begin{equation}
	\frac{F^{cab}F^{abc}}{F^{acb}}=\frac{R^{ac}R^{bc}}{R^{ab,c}}.
	\label{}
\end{equation}
Now we can compute the invariant in Eq.~\eqref{eqn:type3-inv}: 
\begin{equation}
	\begin{split}
		i_\mb{g}\omega(\mb{h,k})&=\frac{\omega(\mb{g,h,k})\omega(\mb{h,k,g})}{\omega(\mb{h,g,k})}\\
	&=\frac{F^{\cohosub{v}(\mb{g}), \cohosub{v}(\mb{h}), \cohosub{v}(\mb{k})}F^{\cohosub{v}(\mb{h}), \cohosub{v}(\mb{k}), \cohosub{v}(\mb{g})}}{F^{\cohosub{v}(\mb{h}), \cohosub{v}(\mb{g}), \cohosub{v}(\mb{k})}}\\
	&=\frac{R^{\cohosub{v}(\mb{h}),\cohosub{v}(\mb{g})}R^{\cohosub{v}(\mb{k}),\cohosub{v}(\mb{g})}}{R^{\cohosub{v}(\mb{hk}),\cohosub{v}(\mb{g})}},
	\end{split}
	\label{}
\end{equation}
and a similar expression for $i_\mb{g}\omega(\mb{k,h})$. Since $\mb{hk}=\mb{kh}$, we find $i_\mb{g}\omega(\mb{h,k})=i_\mb{g}\omega(\mb{k,h})$, contradicting the indicator in Eq.~\eqref{eqn:type3-inv}.

\subsection{Projective symmetries without Kac-Moody algebra}
\label{sec:proj_no_KM}
Here we discuss two examples of symmetry-enriched RCFTs without Kac-Moody algebras~\cite{HampapuraJHEP2016, BaeJHEP2018b}, where the chiral primaries nevertheless transform projectively under the symmetry. Both of them are closely related to the famous Monster CFT, which only has the identity chiral primary, but the chiral algebra has the largest sporadic group, the Monster, as its symmetry. 

The first example is the so-called baby Monster CFT~\cite{Hoehn}. The chiral CFT has $c_L=\frac{47}{2}$ and three chiral primaries with spins $0, \frac{3}{2}, \frac{31}{16}$, so the corresponding MTC is just the $\ol{\mathrm{Ising}}$ MTC (or equivalently Spin(15)$_1$). In fact, it is ``dual'' to the Ising CFT, in the sense that their characters can be combined to give the character of the Monster CFT.  
The global symmetry of the chiral algebra is the baby Monster group (the second largest sporadic group), and the spin $\frac{31}{16}$ chiral primary operator transforms as a projective representation of dimension 96256, see Refs.~\cite{BaeJHEP2018b, LinShaoMonster}.

Another related example is a chiral CFT with $c=\frac{116}{5}$~\cite{Hohn2012}, which is ``dual'' to the three-state Potts CFT. This theory has the largest Fischer group $\mathrm{Fi}_{24}$ as the automorphism group of the chiral algebra. The $h=\frac{4}{3}$ chiral primary transforms as a complex projective representation of $\mathrm{Fi}_{24}$ of dimension 783, see Ref.~\cite{BaeJHEP2018b}.

\subsection{Wess-Zumino-Witten CFTs}
\label{sec:wzw}

Let us review the basics of WZW CFTs. Let $\mathfrak{g}$ be a simple Lie algebra and $k>0$ an integer. Primaries in a $\mathfrak{g}_k$ CFT are labeled by Dynkin labels $\bm{\lambda}=[\lambda_1,\lambda_2,\dots, \lambda_r]$ where $r$ is the rank. They satisfy
\begin{equation}
	0\leq \sum_{i=1}^r a_i^\lor \lambda_i\leq k.
	\label{}
\end{equation}
Here $a_i^\lor$ are the comarks.  The conformal weight of a primary is then given by
\begin{equation}
	h_{\bm{\lambda}}=\frac{(\bm{\lambda}, \bm{\lambda}+2\bm{\rho})}{2(k+h^\lor)},
	\label{}
\end{equation}
where $\bm{\rho}=(1,1,\dots,1)$ is the Weyl vector and $h^\lor$ is the dual Coxeter number.

We only consider diagonal WZW theories in this section. Recall that the continuous symmetry of the $\mathfrak{g}_k$ chiral algebra is
\begin{equation}
	\mathrm{Inn}(\cal{V})=\cal{G}_L/Z(\cal{G}_L)\times \cal{G}_R/Z(\cal{G}_R).
	\label{}
\end{equation}
It is well-known that if the global symmetry group $G$ maps entirely into $\cal{G}_L/Z(\cal{G}_L)$, i.e. acting only on the left-moving fields, there is possibly a chiral anomaly. In this case, though, the naive local operator spectrum may not represent $G$ faithfully. For example, if we identify $G$ with $\cal{G}_L/Z(\cal{G}_L)$, only linear irreps of $G$ can occur in the spectrum. In other words, the original $\cal{G}_k$ theory needs to be ``truncated'', and the proper way to do this is to extend the chiral algebra by simple currents that canonically correspond to $Z(\cal{G})$~\cite{Moore89c}. In the MTC language, the corresponding Abelian anyons form a condensable algebra. This is possible only for $k$'s that satisfy certain divisibility condition. For example, for $\cal{G}=\mathrm{SU}(n)$, $k$ has to be a multiple of $n$ ($2n$) for odd (even) $n$.  The result can be viewed as WZW theory for the non-simply-connected group $\cal{G}/Z(\cal{G})$.

%\begin{equation}
%	(\bm{\lambda}, \bm{\lambda}+2\bm{\rho})=\sum_{i,j}(\lambda_i+2\rho_i)G(\mathfrak{g})_{ij}\lambda_j. 
%\end{equation}
%$G(\mathfrak{g})$ is the inverse Cartan matrix of the Lie algebra $\mathfrak{g}$.

\subsubsection{Center symmetries}
In this section we consider center symmetries in WZW CFTs. We enumerate simple currents in all cases, and also give the corresponding Abelian MTC, using the notation in Ref.~\cite{Bonderson07b}.
Most of the results were already derived in Ref.~\cite{NumasawaJHEP}, except for the $D_{2n}$ series. 
\newline
%We also find conditions for anomaly-free center symmetries~\cite{FelderCMP1988, Ahn1989}

\textbf{$A_{n-1}$ series with $n\geq 2$}, SU$(n)_k$: the center is $\Z_n$. The Dynkin labels for simple currents have one nonzero entry $k$ and all others $0$.  They can be generated from $(0,0,\dots, k)$ by fusion, and the corresponding rep. is obtained from the tensor product of $k$ copies of the fundamental representation $(0,0,\dots,1)$. The spin of the primary is
\begin{equation}
	h_{(0,0,\dots,k)}=\frac{(n-1)k}{2n}.
	\label{}
\end{equation}
In the MTC, this Abelian subcategory is denoted $\Z_n^{(k(n-1)/2)}$. For odd $n$, the F symbols are trivial so the center symmetry is non-anomalous. For even $n$, let us denote the simple objects by $a=0,1,\dots,n-1$,  corresponding to Dynkin label $\lambda_j=\delta_{j,r+1-a}$. Then the F symbols are $F^{abc}=(-1)^{ka\frac{b+c-[b+c]_n}{n}}$.
Thus if $k$ is even the center symmetry is also anomaly-free. 
\newline

\textbf{$B_n$ series with $n\geq 2$}, Spin$(2n+1)_k$: the center is $\Z_2$. The only simple current $(k,0,\dots,0)$ has spin $k/2$.
The Abelian subcategory is $\Z_2^{(k)}$ with trivial F symbols. Therefore the center symmetry is anomaly-free. 
\newline

\textbf{$C_n$ series}, USp$(2n)_k$~\footnote{We use the convention $\mathrm{USp}(2)=\mathrm{SU}(2)$.}: the center is $\Z_2$, with the only simple current given by $(0,0,\dots,k)$, with the spin $\frac{nk}{4}$. Thus the center symmetry is anomaly-free for $nk$ even.
\newline

\textbf{$D_{n}$ series with $n\geq 4$}, Spin$(2n)_k$:  the center group is $\Z_2\times\Z_2$ for even $n$ and $\Z_4$ for odd $n$. The nontrivial simple currents are $(0,0,\dots,k), (0,0,\dots, k,0)$ and $(k,0,\dots,0)$, whose spins are given by
\begin{align}
	h_{(0,\dots,0,k)}= 
	h_{(0,\dots,k,0)}=\frac{kn}{8}, && h_{(k,0,\dots,0)}=\frac{k}{2}.
	\label{}
\end{align}
We remark that $(0,\dots,0,1)$ and $(0,\dots,1,0)$ are spinor representations, and $(1,0,\dots,0)$ is the vector representation.

\begin{itemize}
	\item For $n$ odd, the corresponding Abelian category is $\Z_4^{(kn/2)}$.
	\item For $n$ even and $k$ odd, the Abelian category is isomorphic to Spin$(2nk)_1$. More specifically,
\begin{equation}
	\begin{split}
	n\equiv 0\,(\text{mod 8}),\,&	\mathrm{D}(\Z_2), \\
		n \equiv \pm 2\,(\text{mod 8}), \, &\Z_2^{(kn/4)}\times \Z_2^{(kn/4)},\\
		n\equiv 4\,(\text{mod 8}), \,&\mathrm{Spin}(8)_1.
	\end{split}
	\label{}
\end{equation}
	\item For $n$ even and $k$ even, the Abelian category is $\Z_2^{(\frac{kn}{4})}\times\Z_2^{(\frac{kn}{4})}$, with trivial F symbols.
\end{itemize}

$E_6$: the center is $\Z_3$. The simple currents are $(k,0,\dots,0)$ and $(0,\dots,k,0)$, and both have spin $\frac{2k}{3}$. The Abelian subcategory is $\Z_3^{(k)}$.
\newline

$E_7$: the center is $\Z_2$, and the Abelian subcategory is $\Z_2^{(k/2)}$. Therefore the center symmetry has a $\Z_2$ 't Hooft anomaly for odd $k$.
\newline

The other exceptional Lie groups, $E_8, F_4, G_2$ have trivial centers.

Some of the WZW CFTs either have no simple currents, or the simple currents form an Abelian subcategory with trivial F symbols and braiding. In these cases, the relative anomalies always vanish, provided the symmetry does not act as a nontrivial outer automorphism. We tabulate these theories below in Table~\ref{tab:wzw-no-anomaly}.

\begin{table}
	\centering
\begin{tabular}{|c|c|c|}
	\hline
	WZW model & Condition & Simple currents\\
	\hline
	SU$(n)_k$ & $n|k$ & $\Z_n^{(0)}$\\
	\hline
	Spin$(2n+1)_k$ & & $\Z_2^{(1)}$\\
	\hline
	Spin$(2n)_k$ & $n\text{ odd}, 4|k$ & $\Z_4^{(\frac{kn}{2})}$\\
	\hline
	Spin$(2n)_k$ & $n,k\text{ even}$ & $\Z_2^{(\frac{kn}{4})}\times\Z_2^{(\frac{kn}{4})}$\\
	\hline
	USp$(2n)_k$ & $nk$ even & $\Z_2^{(\frac{nk}{2})}$\\
	\hline
	$(E_6)_k$ & $3|k$ &  $\Z_3^{(0)}$ \\
	\hline
	$(E_7)_k$ & $k$ even & $\Z_2^{(k/2)}$\\
	\hline
	$(E_8)_{k\neq 2}, (F_4)_k, (G_2)_k$ & & $\Z_1$\\
	\hline
	$(E_8)_2$ & &  $\Z_2^{(1)}$\\
	\hline
\end{tabular}
\caption{WZW CFTs in which simple currents have trivial braiding. }
\label{tab:wzw-no-anomaly}
\end{table}

\subsection{Lieb-Schultz-Mattis anomaly}
We now apply the anomaly formula to translation-invariant 1D spin chains which satisfy Lieb-Schultz-Mattis-type (LSM) theorems~\cite{LSM}. It has been understood now that the LSM theorems follow from mixed anomalies between lattice translation and internal symmetry groups~\cite{ChengPRX2016, JianPRB2018, HuangPRB2017, MetlitskiPRB2018}. 

Consider an internal symmetry group $G$. Each site of the spin chain transforms as a projective representation of $G$, labeled by a class $\nu$ in $\H^2[G, \U]$. The spin chain also has translation symmetry $\Z$ generated by a unit translation $T_x$. The LSM anomaly can be understood by treating the translation $\Z$ as an internal symmetry. Intuitively, inserting a unit translation flux is the same as increasing the number of sites by one. Thus the anomaly implies that a unit translation flux transforms as the $[\nu]$ projective representation under $G$. Technically, the 2-cocycle $[\nu]$ can be extracted from the anomaly 3-cocycle with the help of slant product:
\begin{equation}
	[\nu]=[i_{T_x}\omega]\big|_G,
	\label{}
\end{equation}
where $\big|_G$ means restriction to $G$.

Suppose that the CFT is diagonal and symmetries do not permute primaries (i.e $\rho=\mb{1}$). Using the definition of slant product one finds
\begin{equation}
	\nu(\mb{g,h})\simeq M_{\cohosub{v}(T_x), \cohosub{w}(\mb{g,h})}{M_{\cohosub{v}(\mb{h}), \cohosub{b}(\mb{g},T_x)}}.
	\label{}
\end{equation}
Here $\simeq$ means equivalence as cohomology classes, and $\coho{b}(\mb{h,k})=\coho{w}(\mb{h,k})\times\ol{\coho{w}(\mb{k,h})}$. One can show that $\coho{b}(\cdot, T_x)$ defines a homomorphism from $G$ to $\cal{A}$. 

Let us first consider the case where $G$ is a continuous, connected group, e.g. $G=\mathrm{PSU}(N)$. In this case, $\coho{b}$ is always trivial, so only the first factor $ M_{\cohosub{v}(T_x), \cohosub{w}(\mb{g,h})}$ contributes. Observe that the relation now says $\coho{v}(T_x)$ transforms under $G$ according to the projective class $[\nu]$, which agrees with the heuristic argument since the Verlinde line corresponding to $\coho{v}(T_x)$ generates the translation symmetry. 

Suppose $G=\mathrm{PSU}(N)$ and since $\H^2[\mathrm{PSU}(N), \U]=\Z_N$, we consider each site transforming as the $m$-th class in $\Z_N$ (e.g. the symmetric rank-$m$ tensor representation). The chiral algebra of the CFT must contain an $\mathfrak{su}(N)$ Kac-Moody algebra, so it is natural to consider SU$(N)_k$ CFT with $\mathrm{Inn}(\cal{V})=\mathrm{PSU}(N)$.  We have found that the Abelian anyons in the chiral MTC form a $\Z_N$ group, whose $j$-th element transforms as the rank-$kj$ tensor. The anomaly matching condition then requires that there exists $\coho{v}(T_x)=j$ such that
\begin{equation}
	kj\equiv m\,(\text{mod }N).
	\label{}
\end{equation}
The equation is solvable if and only if $(k, N)|m$. For example, there are always solutions $k=1$ or $k=m$. In fact, it was known that $\mathrm{SU}(N)_k$ is realized in a generalized Heisenberg chain where each site transforms as the symmetric rank-$k$ tensor representation~\cite{Sutherland1975, Andrei1984, Johannesson1986, Alcaraz1989, DMRGSUN}.  When $m\neq 0$, the ``minimal'' theory is always $k=1$, the SU$(N)_1$ CFT, and $\coho{v}(T_x)=[m]$. Our results agree with those in Ref.~\cite{Yao2018}, which were obtained using different methods.

Essentially the same LSM anomaly exists if PSU$(N)$ is broken down to the $\Z_N\times\Z_N$ subgroup.  Denote elements of the symmetry group additively as $\Z_N\times\Z_N\times\Z$ by $\mb{a}=(a_1, a_2, a_3)$ where $a,b\in \{0,1,\dots, N-1\}$ and $a_3\in\Z$. The anomaly 3-cocycle takes the following form:
\begin{equation}
	\omega(\mb{a}, \mb{b}, \mb{c}) = e^{\frac{2\pi i p}{N}a_1b_2c_3}.
	\label{}
\end{equation}
Here $p\in \Z/N\Z$.   

Let us work this out explicitly for SU$(N)_k$ CFTs, whose MTC has a $\Z_N^{(k(N-1)/2)}$ Abelian subcategory. For simplicity we assume $N$ is odd, so the F symbols are all $1$. A sufficient condition for anomaly matching is:
\begin{equation}
	k\coho{v}(\mb{a})\coho{w}(\mb{b,c}) \equiv pa_1b_2c_3\, (\text{mod }N).
	\label{}
\end{equation}
Set $\coho{v}(\mb{a})=xa_1, \coho{w}(\mb{b,c})=b_2c_3$, then $kx=p\,\text{mod }N$, which is solvable as long as $(k, N)|p$. From the projective representation carried by the $(0,0,\dots, k)$, we can see that this theory is the one obtained from breaking PSU$(N)$ down to $\Z_N\times\Z_N$.  Essentially the same is true for even $N$.
Notice that all results so far stay formally the same even if the translation symmetry group $\Z$ is further broken down to $\Z_N$. This is an example of the type-III cocycle discussed in Sec.~\ref{sec:Gcentral}. We remark that the cohomology class is invariant under arbitrary permutations of the three $\Z_N$ subgroups. For example, 
 we may set $\coho{v}(\mb{a})=a_2, \coho{w}(\mb{b,c})=b_1c_3$, and the anomaly condition is satisfied as well. 

 SU$(N)_1$ can be viewed as a Luttinger liquid with $(N-1)$ components, thus admitting marginal deformations by tuning Luttinger parameters. We are not aware of any RCFTs with $c<N-1$ that can saturate this anomaly. It was recently conjectured that any CFTs with the $\Z_N\times\Z_N$ LSM anomaly must have a minimal central charge $N-1$~\cite{Alvirad2019, NingSPTedge}. This is obviously true for the PSU$(N)$ case, as the chiral algebra must contain a $\mathfrak{su}(n)$ Kac-Moody algebra with a minimal central charge $N-1$, but it is much less obvious that the lower bound still holds for $G=\Z_N\times\Z_N$. 

 This analysis can be easily generalized to other Lie groups. Suppose $G=\cal{G}/Z(\cal{G})$ where $\cal{G}$ is a simple Lie group. Since the CFT should have $G$ symmetry, $\cal{G}_k$ are the most natural candidates. Clearly $G$ must be identified as the diagonal subgroup of $\mathrm{Inn}(\cal{G}_k)$. Since the translation commutes with $G$, it must be realized as a central symmetry, generated by an invertible Verlinde line $\coho{v}(T_x)$ in $\cal{A}\simeq Z(\cal{G})$. In fact, in this case the relative anomaly becomes absolute (the ``reference'' theory is the one with trivial translation action). The LSM anomaly requires that $\coho{v}(T_x)$ carries the same projective representation as the site Hilbert space, which can be easily checked using the results in Sec.~\ref{sec:wzw}. In particular, according to Table~\ref{tab:wzw-no-anomaly} there are two families of groups, $G=\mathrm{SO}(2n+1)$ or $\mathrm{PSp}(2n)\equiv\mathrm{USp}(4n)/\Z_2$, whose relative anomalies vanish regardless of the value of $k$. Both families allow $\Z_2$-classified projective representations (e.g. $2^n$-dimensional spinor representations for $\mathrm{Spin}(2n+1)$, and $4n$-dimensional fundamental representations for $\mathrm{USp}(4n)$), and therefore the corresponding spin chain can not have $\cal{G}_k$ as its low-energy theory without breaking the symmetry, as observed in Ref. \cite{metlitski}.

 \section{Discussions}
 In this paper we have extensively analyzed symmetry actions on local operators in a (1+1)d CFT. This is not the entire story for symmetry-enriched CFTs, though, as non-local operators, such as symmetry defect operators, are also important. In fact, the 't Hooft anomaly itself is a fundamental property of symmetry defect operators. In Ref.~\cite{Verrensen2019}, it was shown that for certain symmetry-enriched CFTs, given the symmetry actions on local operators there are additional phases distinguished by charges carried by symmetry defects, which can lead to robust edge modes. Roughly speaking, one may stack (1+1)d gapped SPT states onto the CFT to ``toggle'' between these phases. Whether stacking an SPT state leads to a distinct phase or not depends strongly on the symmetry property of topological defect operators. We will investigate these issues in future work.

In this work we exclusively considered unitary symmetries. It is not clear whether our approach can be generalized to anti-unitary symmetry, as the ``strip'' picture naively breaks time-reversal symmetry. Developing the theory of anti-unitary symmetry-enriched CFTs is an important direction for future research.
Another limitation of the work is that our results only apply to bosonic systems. The classification of (2+1)d fermionic SPT phases has been established recently~\cite{fspt, Bhardwaj2017fspt}. On the CFT side, it is important to extend the results to fermionic CFTs. 

Recently non-invertible anomalies have also been studied in (1+1)d CFTs~\cite{ Bhardwaj2017, ChangTDL, Vanhove2018, JiPRR2019, WangYF2019, KongCFT2019a, KongCFT2019b}. As a special case, any CFT with a global unitary symmetry $G$, anomalous or not, can be coupled to a (2+1)d $G$ gauge theory as an edge CFT, in which case the operator content is very different from a true (1+1)d theory. Many results in this work can be effortlessly translated to this context. For example, it easily follows from the discussions in Sec.~\ref{sec:mm} that edge CFTs of any twisted Dijkgraaf-Witten gauge theory must have $c\geq 1$~\cite{JiPRR2019} if the gauge symmetry is not broken on the edge.  

\section{Acknowledgement}
MC is grateful for Chao-Ming Jian and Shang-Qiang Ning for enlightening conversations, Zhenghan Wang, Po-Shen Hsin and Liang Kong for helpful correspondence. MC would like to especially thank Wenjie Ji, Max Metlitski and Shu-Heng Shao for extensive discussions on (1+1)d CFTs. DW would like to thank Dave Aasen and Corey Jones for related discussions about symmetric gapped boundaries of (2+1)d SETs.  The work is supported by NSF under award number DMR-1846109 and the Alfred P. Sloan foundation.  

\appendix

\tikzset{middlearrow/.style={
        decoration={markings,
            mark= at position 0.55 with {\arrow{#1}} ,
        },
        postaction={decorate}
    }
}

\tikzset{
  % style to apply some styles to each segment of a path
  on each segment/.style={
    decorate,
    decoration={
      show path construction,
      moveto code={},
      lineto code={
        \path [#1]
        (\tikzinputsegmentfirst) -- (\tikzinputsegmentlast);
      },
      curveto code={
        \path [#1] (\tikzinputsegmentfirst)
        .. controls
        (\tikzinputsegmentsupporta) and (\tikzinputsegmentsupportb)
        ..
        (\tikzinputsegmentlast);
      },
      closepath code={
        \path [#1]
        (\tikzinputsegmentfirst) -- (\tikzinputsegmentlast);
      },
    },
  },
  % style to add an arrow in the middle of a path
  mid arrow/.style={postaction={decorate,decoration={
        markings,
        mark=at position .7 with {\arrow[#1]{Stealth}}
      }}},
  smid arrow/.style={postaction={decorate,decoration={
        markings,
        mark=at position .7 with {\arrow[#1]{stealth}}
      }}},
}

\tikzset{anyon/.style={line width=1pt,draw=black,postaction={on each segment={smid arrow=black}}}}
\tikzset{dw/.style={line width=1pt,draw=black,postaction={on each segment={smid arrow=black}}, dashed}}
\tikzset{vacuum/.style={line width=1pt, draw=black, loosely dotted}}
\tikzset{UFCBackground/.style={draw=black!10!white!80,fill=black!10!white!80}}

\newcommand{\drawtube}[4]{
\begin{tikzpicture}[scale=.1]
\def\dx{-1};
\def\ddx{1.5};
\def\dy{.5};
\def\p{5.5};
\filldraw[UFCBackground](0,0)--(20,0)--(20,20)--(0,20)--(0,0);
\draw[anyon](13,9)--(20,9) node[pos=0.55,above] {$#3$};
\draw[anyon](7,11)--(13,9) node[pos=0.45,below] {$#2$};
\draw[anyon](0,11)--(7,11) node[pos=0.55,above] {$#1$};
\draw[anyon](10,0)--(13,9)  node[pos=0.3,right] {$#4$};
\draw[anyon](7,11)--(10,20) node[pos=0.55,right] {$#4$};
\draw[draw=black!29,dashed](0,0)--(20,0) (20,20)--(0,20);
\filldraw[draw=black!29,fill=black!29](\p,0)--(\p+\dx,-\dy)--(\p+\dx,\dy)--(\p,0);
\filldraw[draw=black!29,fill=black!29](\p,20)--(\p+\dx,20-\dy)--(\p+\dx,20+\dy)--(\p,20);
\filldraw[draw=black!29,fill=black!29](\p+\ddx,0)--(\p+\dx+\ddx,-\dy)--(\p+\dx+\ddx,\dy)--(\p+\ddx,0);
\filldraw[draw=black!29,fill=black!29](\p+\ddx,20)--(\p+\dx+\ddx,20-\dy)--(\p+\dx+\ddx,20+\dy)--(\p+\ddx,20);
	\end{tikzpicture}
	}
	
\newcommand{\drawsettube}[4]{
\begin{tikzpicture}[scale=.1]
\def\dx{-1};
\def\ddx{1.5};
\def\dy{.5};
\def\p{5.5};
\filldraw[UFCBackground](0,0)--(20,0)--(20,20)--(0,20)--(0,0);
\draw[anyon](13,9)--(20,9) node[pos=0.55,above] {$#3$};
\draw[anyon](7,11)--(13,9) node[pos=0.25,below] {$#2$};
\draw[anyon](0,11)--(7,11) node[pos=0.55,above] {$#1$};
\draw[anyon](10,0)--(13,9)  node[pos=0.3,right] {$#4$};
\draw[anyon](7,11)--(10,20) node[pos=0.55,right] {$#4$};
\draw[draw=black!29,dashed](0,0)--(20,0) (20,20)--(0,20);
\filldraw[draw=black!29,fill=black!29](\p,0)--(\p+\dx,-\dy)--(\p+\dx,\dy)--(\p,0);
\filldraw[draw=black!29,fill=black!29](\p,20)--(\p+\dx,20-\dy)--(\p+\dx,20+\dy)--(\p,20);
\filldraw[draw=black!29,fill=black!29](\p+\ddx,0)--(\p+\dx+\ddx,-\dy)--(\p+\dx+\ddx,\dy)--(\p+\ddx,0);
\filldraw[draw=black!29,fill=black!29](\p+\ddx,20)--(\p+\dx+\ddx,20-\dy)--(\p+\dx+\ddx,20+\dy)--(\p+\ddx,20);
	\end{tikzpicture}
	}
	
\newcommand{\drawisingtube}[4]{
\begin{tikzpicture}[scale=.1]
\def\dx{-1};
\def\ddx{1.5};
\def\dy{.5};
\def\p{5.5};
\filldraw[UFCBackground](0,0)--(20,0)--(20,20)--(0,20)--(0,0);
\draw[#3](20,9)--(13,9);
\draw[#2](13,9)--(7,11);
\draw[#1](7,11)--(0,11);
\draw[#4](13,9)--(10,0);
\draw[#4](10,20)--(7,11);
\draw[draw=black!29,dashed](0,0)--(20,0) (20,20)--(0,20);
\filldraw[draw=black!29,fill=black!29](\p,0)--(\p+\dx,-\dy)--(\p+\dx,\dy)--(\p,0);
\filldraw[draw=black!29,fill=black!29](\p,20)--(\p+\dx,20-\dy)--(\p+\dx,20+\dy)--(\p,20);
\filldraw[draw=black!29,fill=black!29](\p+\ddx,0)--(\p+\dx+\ddx,-\dy)--(\p+\dx+\ddx,\dy)--(\p+\ddx,0);
\filldraw[draw=black!29,fill=black!29](\p+\ddx,20)--(\p+\dx+\ddx,20-\dy)--(\p+\dx+\ddx,20+\dy)--(\p+\ddx,20);
	\end{tikzpicture}
	}
	
\section{Review of Unitary MTCs}
\label{sec:mtc-review}
The topologically non-trivial quasiparticles of a (2+1)D topologically ordered state are equivalently referred to
as anyons, topological charges, and quasiparticles. In the category theory terminology, they correspond
to isomorphism classes of simple objects  in a unitary MTC (UMTC). 

A UMTC $\mathcal{C}$ contains splitting spaces $V_{c}^{ab}$, and their dual fusion spaces, $V_{ab}^c$,
where $a,b,c \in \mathcal{C}$ are the anyons. These spaces have dimension 
$\text{dim } V_{c}^{ab} = \text{dim } V_{ab}^c = N_{ab}^c$, where $N_{ab}^c$ are referred
to as the fusion rules. They are depicted graphically as: 
\begin{equation}
\left( d_{c} / d_{a}d_{b} \right) ^{1/4}
\pspicture[shift=-0.6](-0.1,-0.2)(1.5,-1.2)
  \small
  \psset{linewidth=0.9pt,linecolor=black,arrowscale=1.5,arrowinset=0.15}
  \psline{-<}(0.7,0)(0.7,-0.35)
  \psline(0.7,0)(0.7,-0.55)
  \psline(0.7,-0.55) (0.25,-1)
  \psline{-<}(0.7,-0.55)(0.35,-0.9)
  \psline(0.7,-0.55) (1.15,-1)	
  \psline{-<}(0.7,-0.55)(1.05,-0.9)
  \rput[tl]{0}(0.4,0){$c$}
  \rput[br]{0}(1.4,-0.95){$b$}
  \rput[bl]{0}(0,-0.95){$a$}
 \scriptsize
  \rput[bl]{0}(0.85,-0.5){$\mu$}
  \endpspicture
=\left\langle a,b;c,\mu \right| \in
V_{ab}^{c} ,
\label{eq:bra}
\end{equation}
\begin{equation}
\left( d_{c} / d_{a}d_{b}\right) ^{1/4}
\pspicture[shift=-0.65](-0.1,-0.2)(1.5,1.2)
  \small
  \psset{linewidth=0.9pt,linecolor=black,arrowscale=1.5,arrowinset=0.15}
  \psline{->}(0.7,0)(0.7,0.45)
  \psline(0.7,0)(0.7,0.55)
  \psline(0.7,0.55) (0.25,1)
  \psline{->}(0.7,0.55)(0.3,0.95)
  \psline(0.7,0.55) (1.15,1)	
  \psline{->}(0.7,0.55)(1.1,0.95)
  \rput[bl]{0}(0.4,0){$c$}
  \rput[br]{0}(1.4,0.8){$b$}
  \rput[bl]{0}(0,0.8){$a$}
 \scriptsize
  \rput[bl]{0}(0.85,0.35){$\mu$}
  \endpspicture
=\left| a,b;c,\mu \right\rangle \in
V_{c}^{ab},
\label{eq:ket}
\end{equation}
where $\mu=1,\ldots ,N_{ab}^{c}$, $d_a$ is the quantum dimension of $a$, 
and the factors $\left(\frac{d_c}{d_a d_b}\right)^{1/4}$ are a normalization convention for the diagrams. 

We denote $\bar{a}$ as the topological charge conjugate of $a$, for which
$N_{a \bar{a}}^1 = 1$, i.e.
\begin{align}
a \times \bar{a} = 1 +\cdots
\end{align}
Here $1$ refers to the identity particle, i.e. the vacuum topological sector, which physically describes all 
local, topologically trivial excitations. 

The $F$-symbols are defined as the following basis transformation between the splitting
spaces of $4$ anyons:
\begin{equation}
  \pspicture[shift=-1.0](0,-0.45)(1.8,1.8)
  \small
 %%%% Upper fan:
  \psset{linewidth=0.9pt,linecolor=black,arrowscale=1.5,arrowinset=0.15}
  \psline(0.2,1.5)(1,0.5)
  \psline(1,0.5)(1,0)
  \psline(1.8,1.5) (1,0.5)
  \psline(0.6,1) (1,1.5)
   \psline{->}(0.6,1)(0.3,1.375)
   \psline{->}(0.6,1)(0.9,1.375)
   \psline{->}(1,0.5)(1.7,1.375)
   \psline{->}(1,0.5)(0.7,0.875)
   \psline{->}(1,0)(1,0.375)
   %%% Labels:
   \rput[bl]{0}(0.05,1.6){$a$}
   \rput[bl]{0}(0.95,1.6){$b$}
   \rput[bl]{0}(1.75,1.6){${c}$}
   \rput[bl]{0}(0.5,0.5){$e$}
   \rput[bl]{0}(0.9,-0.3){$d$}
 \scriptsize
   \rput[bl]{0}(0.3,0.8){$\alpha$}
   \rput[bl]{0}(0.7,0.25){$\beta$}
  \endpspicture
%%%
= \sum_{f,\mu,\nu} \left[F_d^{abc}\right]_{(e,\alpha,\beta)(f,\mu,\nu)}
%%%
 \pspicture[shift=-1.0](0,-0.45)(1.8,1.8)
  \small
 %%%% Upper fan:
  \psset{linewidth=0.9pt,linecolor=black,arrowscale=1.5,arrowinset=0.15}
  \psline(0.2,1.5)(1,0.5)
  \psline(1,0.5)(1,0)
  \psline(1.8,1.5) (1,0.5)
  \psline(1.4,1) (1,1.5)
   \psline{->}(0.6,1)(0.3,1.375)
   \psline{->}(1.4,1)(1.1,1.375)
   \psline{->}(1,0.5)(1.7,1.375)
   \psline{->}(1,0.5)(1.3,0.875)
   \psline{->}(1,0)(1,0.375)
   %%% Labels:
   \rput[bl]{0}(0.05,1.6){$a$}
   \rput[bl]{0}(0.95,1.6){$b$}
   \rput[bl]{0}(1.75,1.6){${c}$}
   \rput[bl]{0}(1.25,0.45){$f$}
   \rput[bl]{0}(0.9,-0.3){$d$}
 \scriptsize
   \rput[bl]{0}(1.5,0.8){$\mu$}
   \rput[bl]{0}(0.7,0.25){$\nu$}
  \endpspicture
.
\end{equation}
To describe topological phases, these are required to be unitary transformations.

The $R$-symbols define the braiding properties of the anyons, and are defined via the the following
diagram:
\begin{equation}
\pspicture[shift=-0.65](-0.1,-0.2)(1.5,1.2)
  \small
  \psset{linewidth=0.9pt,linecolor=black,arrowscale=1.5,arrowinset=0.15}
  \psline{->}(0.7,0)(0.7,0.43)
  \psline(0.7,0)(0.7,0.5)
% \psarc(0.9,0.846410){0.4}{120}{240}
% \psarc(0.5,0.846410){0.4}{-60}{40}
 \psarc(0.8,0.6732051){0.2}{120}{240}
 \psarc(0.6,0.6732051){0.2}{-60}{35}
  \psline (0.6134,0.896410)(0.267,1.09641)
  \psline{->}(0.6134,0.896410)(0.35359,1.04641)
  \psline(0.7,0.846410) (1.1330,1.096410)	
  \psline{->}(0.7,0.846410)(1.04641,1.04641)
  \rput[bl]{0}(0.4,0){$c$}
  \rput[br]{0}(1.35,0.85){$b$}
  \rput[bl]{0}(0.05,0.85){$a$}
 \scriptsize
  \rput[bl]{0}(0.82,0.35){$\mu$}
  \endpspicture
=\sum\limits_{\nu }\left[ R_{c}^{ab}\right] _{\mu \nu}
\pspicture[shift=-0.65](-0.1,-0.2)(1.5,1.2)
  \small
  \psset{linewidth=0.9pt,linecolor=black,arrowscale=1.5,arrowinset=0.15}
  \psline{->}(0.7,0)(0.7,0.45)
  \psline(0.7,0)(0.7,0.55)
  \psline(0.7,0.55) (0.25,1)
  \psline{->}(0.7,0.55)(0.3,0.95)
  \psline(0.7,0.55) (1.15,1)	
  \psline{->}(0.7,0.55)(1.1,0.95)
  \rput[bl]{0}(0.4,0){$c$}
  \rput[br]{0}(1.4,0.8){$b$}
  \rput[bl]{0}(0,0.8){$a$}
 \scriptsize
  \rput[bl]{0}(0.82,0.37){$\nu$}
  \endpspicture
  .
\end{equation}

Under a basis transformation, $\Gamma^{ab}_c : V^{ab}_c \rightarrow V^{ab}_c$, the $F$ and $R$ symbols change:
\begin{align}
  F^{abc}_d &\rightarrow \tilde{F}^{abc}_d = \Gamma^{ab}_e \Gamma^{ec}_d F^{abc}_d [\Gamma^{bc}_f]^\dagger [\Gamma^{af}_d]^\dagger
  \nonumber \\
  R^{ab}_c & \rightarrow \tilde{R}^{ab}_c = \Gamma^{ba}_c R^{ab}_c [\Gamma^{ab}_c]^\dagger .
  \end{align}
  These basis transformations are referred to as vertex basis gauge transformations. Physical quantities correspond to gauge-invariant combinations
  of the data. 
  
The topological twist $\theta_a = e^{2\pi i h_a}$, with $h_a$ the topological spin, is defined
via the diagram:
\begin{equation}
\theta _{a}=\theta _{\bar{a}}
=\sum\limits_{c,\mu } \frac{d_{c}}{d_{a}}\left[ R_{c}^{aa}\right] _{\mu \mu }
= \frac{1}{d_{a}}
\pspicture[shift=-0.5](-1.3,-0.6)(1.3,0.6)
\small
  \psset{linewidth=0.9pt,linecolor=black,arrowscale=1.5,arrowinset=0.15}
  %%% Arcs
  \psarc[linewidth=0.9pt,linecolor=black] (0.7071,0.0){0.5}{-135}{135}
  \psarc[linewidth=0.9pt,linecolor=black] (-0.7071,0.0){0.5}{45}{315}
  %%% Straight segments
  \psline(-0.3536,0.3536)(0.3536,-0.3536)
  \psline[border=2.3pt](-0.3536,-0.3536)(0.3536,0.3536)
  \psline[border=2.3pt]{->}(-0.3536,-0.3536)(0.0,0.0)
  \rput[bl]{0}(-0.2,-0.5){$a$}
  \endpspicture
.
\end{equation}
Finally, the modular, or topological, $S$-matrix, is defined as
\begin{equation}
S_{ab} =\mathcal{D}^{-1}\sum%
\limits_{c}N_{\bar{a} b}^{c}\frac{\theta _{c}}{\theta _{a}\theta _{b}}d_{c}
=\frac{1}{\mathcal{D}}
\pspicture[shift=-0.4](0.0,0.2)(2.6,1.3)
\small
  \psarc[linewidth=0.9pt,linecolor=black,arrows=<-,arrowscale=1.5,arrowinset=0.15] (1.6,0.7){0.5}{167}{373}
  \psarc[linewidth=0.9pt,linecolor=black,border=3pt,arrows=<-,arrowscale=1.5,arrowinset=0.15] (0.9,0.7){0.5}{167}{373}
  \psarc[linewidth=0.9pt,linecolor=black] (0.9,0.7){0.5}{0}{180}
  \psarc[linewidth=0.9pt,linecolor=black,border=3pt] (1.6,0.7){0.5}{45}{150}
  \psarc[linewidth=0.9pt,linecolor=black] (1.6,0.7){0.5}{0}{50}
  \psarc[linewidth=0.9pt,linecolor=black] (1.6,0.7){0.5}{145}{180}
  \rput[bl]{0}(0.1,0.45){$a$}
  \rput[bl]{0}(0.8,0.45){$b$}
  \endpspicture
,
\label{eqn:mtcs}
\end{equation}
where $\mathcal{D} = \sqrt{\sum_a d_a^2}$.

A quantity that we make extensive use of is the double braid, which is
a phase if either $a$ or $b$ is an Abelian anyon:
\begin{equation}
  \pspicture[shift=-0.6](0.0,-0.05)(1.1,1.45)
  \small
%%%%% Small oval:
  \psarc[linewidth=0.9pt,linecolor=black,border=0pt] (0.8,0.7){0.4}{120}{225}
  \psarc[linewidth=0.9pt,linecolor=black,arrows=<-,arrowscale=1.4,
    arrowinset=0.15] (0.8,0.7){0.4}{165}{225}
  \psarc[linewidth=0.9pt,linecolor=black,border=0pt] (0.4,0.7){0.4}{-60}{45}
  \psarc[linewidth=0.9pt,linecolor=black,arrows=->,arrowscale=1.4,
    arrowinset=0.15] (0.4,0.7){0.4}{-60}{15}
  \psarc[linewidth=0.9pt,linecolor=black,border=0pt]
(0.8,1.39282){0.4}{180}{225}
  \psarc[linewidth=0.9pt,linecolor=black,border=0pt]
(0.4,1.39282){0.4}{-60}{0}
  \psarc[linewidth=0.9pt,linecolor=black,border=0pt]
(0.8,0.00718){0.4}{120}{180}
  \psarc[linewidth=0.9pt,linecolor=black,border=0pt]
(0.4,0.00718){0.4}{0}{45}
%%%%% Lines:
%   \psset{linewidth=0.9pt,linecolor=black,arrowscale=1.5,arrowinset=0.15}
%   \psline(0.6,1.05)(0.6,1.55)
%   \psline{->}(0.6,1.05)(0.6,1.45)
%   \psline(0.6,-0.15)(0.6,0.35)
%   \psline{->}(0.6,-0.15)(0.6,0.25)
%%%%% Labels:
  \rput[bl]{0}(0.1,1.2){$a$}
  \rput[br]{0}(1.06,1.2){$b$}
  \endpspicture
= M_{ab}
\pspicture[shift=-0.6](-0.2,-0.45)(1.0,1.1)
  \small
  \psset{linewidth=0.9pt,linecolor=black,arrowscale=1.5,arrowinset=0.15}
  \psline(0.3,-0.4)(0.3,1)
  \psline{->}(0.3,-0.4)(0.3,0.50)
  \psline(0.7,-0.4)(0.7,1)
  \psline{->}(0.7,-0.4)(0.7,0.50)
  \rput[br]{0}(0.96,0.8){$b$}
  \rput[bl]{0}(0,0.8){$a$}
  \endpspicture
.
\end{equation}

\section{Global symmetry in a UMTC}
\subsection{Topological symmetry and braided auto-equivalence}
An important property of a UMTC $\mathcal{C}$ is the group of ``topological symmetries,'' which are related
to ``braided auto-equivalences'' in the mathematical literature. They are associated with the discrete symmetries of the
emergent TQFT described by $\mathcal{C}$, irrespective of any microscopic global symmetries of a quantum
system in which the TQFT emerges as the long wavelength description. 

The topological symmetries consist of the invertible maps
\begin{align}
\varphi: \mathcal{C} \rightarrow \mathcal{C} .
\end{align}
The different $\varphi$, modulo equivalences known as natural isomorphisms, form a group, which we denote
as Aut$(\mathcal{C})$~\cite{SET}.

The maps $\varphi$ may permute the topological charges:
\begin{align}
\varphi(a) = a' \in \mathcal{C}, 
\end{align}
subject to the constraint that 
\begin{align}
N_{a'b'}^{c'} = N_{ab}^c, &&
S_{a'b'} = S_{ab}, &&
\theta_{a'} = \theta_a.
\end{align}
The maps $\varphi$ have a corresponding action on the $F$- and $R-$ symbols of the theory,
as well as on the fusion and splitting spaces, which we discuss in the subsequent section. 

\subsection{Global symmetry}
\label{globsym}

Let us now suppose that we are interested in a system with a global symmetry group $G$. For example, we may be interested
in a given microscopic Hamiltonian that has a global symmetry group $G$, whose ground state preserves $G$, and whose 
anyonic excitations are algebraically described by $\mathcal{C}$. The global symmetry acts on the topological quasiparticles and the topological state space through the action of a group homomorphism
\begin{align}
[\rho] : G \rightarrow \text{Aut}(\mathcal{C}) . 
\end{align}
We use the notation $[\rho_{\bf g}] \in \text{Aut}(\mathcal{C})$ for a specific element ${\bf g} \in G$. The square
brackets indicate the equivalence class of symmetry maps related by natural isomorphisms (see Ref.~\cite{SET} for more details). $\rho_{\bf g}$ is thus a
representative map of the equivalence class $[\rho_{\bf g}]$. We use the notation
\begin{align}
\,^{\bf g}a \equiv \rho_{\bf g}(a). 
\end{align}

The map $\rho_{\bf g}$ has an action on the fusion/splitting spaces:
\begin{align}
\rho_{\bf g} : V_{ab}^c \rightarrow V_{\,^{\bf g}a \,^{\bf g}b}^{\,^{\bf g}c} .   
\end{align}
In the following we consider theories with one-dimensional fusion/splitting spaces (i.e. $N_{ab}^c=0,1$), so we write this explicitly as
\begin{align}
\rho_{\bf g} |a,b;c\rangle = U_{\bf g}(\,^{\bf g}a ,
  \,^{\bf g}b ; \,^{\bf g}c )|\,^{\bf g} a, \,^{\bf g} b; \,^{\bf g}c\rangle,
\end{align}
where $U_{\bf g}(\,^{\bf g}a , \,^{\bf g}b ; \,^{\bf g}c ) $ is a phase factor (in general it is an $N_{ab}^c \times N_{ab}^c$ matrix).

The $F$ and $R$ symbols also transform under the map $\rho_{\bf g}$. Invariance under the map requires
\begin{widetext}
\begin{equation}
	\begin{split}
		[F^{abc}_{d}]_{ef} &= U_{\bf g}(\,^{\bf g}a, \,^{\bf g}b; \,^{\bf g}e) U_{\bf g}(\,^{\bf g}e, \,^{\bf g}c; \,^{\bf g}d) [F^{\,^{\bf g}a \,^{\bf g}b \,^{\bf g}c }_{\,^{\bf g}d}]_{\,^{\bf g}e \,^{\bf g}f} 
		U^{-1}_{\bf g}(\,^{\bf g}b, \,^{\bf g}c; \,^{\bf g}f) U^{-1}_{\bf g}(\,^{\bf g}a, \,^{\bf g}f; \,^{\bf g}d).
 \\
 R^{ab}_c &= U_{\bf g}(\,^{\bf g}b, \,^{\bf g}a; \,^{\bf g}c)  R^{\,^{\bf g}a \,^{\bf g}b}_{\,^{\bf g}c} U_{\bf g}(\,^{\bf g}a, \,^{\bf g}b; \,^{\bf g}c)^{-1},
\end{split}
\label{eqn:FRsym}
\end{equation}
\end{widetext}
where we have suppressed the additional indices that appear when $N_{ab}^c > 1$.

Now let us consider the action of a symmetry ${\bf g} \in G$ on the full many-body state of the system.
Let $R_{\bf g}$ be the representation of ${\bf g}$ acting on the full Hilbert space of the theory. 
We consider a state $|\{a_j\}_{j=1}^n \rangle$ in the full Hilbert space of the system, 
which consists of $n$ anyons, $a_1, \cdots a_n$, at well-separated locations, which collectively fuse to the identity topological sector. 
Since the ground state is $G$-symmetric, we expect that the symmetry action $R_{\bf g}$ on this state possesses a property
that we refer to as symmetry localization. This is the property that the symmetry action $R_{\bf g}$ decomposes as
\begin{align}
\label{eqn:symloc}
R_{\bf g}  \ket{\{a_j\}} \approx \prod_{j = 1}^n U^{(j)}_{\bf g} \rho_\mb{g} \ket{\{a_j\}}.
\end{align}
Here, $U^{(j)}_{\bf g}$ are unitary operators that have support in a region (of length scale set by the correlation length)
localized to the anyon $a_j$. They satisfy the group multiplication up to projective phases:
\begin{equation}
	{}^\mb{g}U_\mb{h}^{(j)}U_\mb{g}^{(j)}\ket{\{a_j\}}=\eta_{a_j}(\mb{g,h})U_\mb{gh}^{(j)}\ket{\{a_j\}}.
	\label{}
\end{equation}
Here ${}^\mb{g}O=R_\mb{g}OR_\mb{g}^{-1}$.

The map $\rho_\mb{g}$ only depends on the global topological sector of the system -- that is, on the 
precise fusion tree that defines the topological state -- and not on any other details of the state, in contrast to the local operators $U^{(j)}_{\bf g}$.
The $\approx$ means that the equation is true up to corrections that are exponentially small in the size of $U^{(j)}$ and the distance between the anyons,
in units of the correlation length. 

To write down the explicit form of $\rho_\mb{g}$, let us specify a particular fusion tree.  Suppose the intermediate fusion channels are $c_{1}, c_{2}, \dots, c_{n-1}$ such that
\begin{align}
	N_{a_1,a_2}^{c_{1}}>0, && N_{c_{1},a_3}^{c_{2}}>0, && \dots, && N_{c_{n-1},a_n}^0>0.
	\label{}
\end{align}
We now include the $c_j$'s in the label of the state, so write $\ket{\{a_j\};\{c_j\}}$.
Now $\rho_\mb{g}$ acting on topological state space is given by
\begin{widetext}
	\begin{equation}
	\rho_\mb{g}\ket{\{a_j\};\{c_j\}}=U_\mb{g}( {}^\mb{g}a_1, {}^\mb{g}a_2; {}^\mb{g}c_1)U_\mb{g}( {}^\mb{g}c_1, {}^\mb{g}a_3; {}^\mb{g}c_2)\cdots U_\mb{g}( {}^\mb{g}c_{n-1}, {}^\mb{g}a_n; 0)\ket{\{{}^\mb{g}a_j\};\{ {}^\mb{g}c_j\}}.
	\label{eqn:rho_g}
\end{equation}
\end{widetext}

Starting from Eq.~\eqref{eqn:symloc}, one finds that $U$ and $\eta$ further satisfy the following two consistency conditions~\cite{SET}:
\begin{equation}
	\eta_{\rho_{\mb{g}}^{-1}(a)}(\mb{h,k})  \eta_{a}(\mb{g,hk}) = \eta_{a}(\mb{gh,k})  \eta_{a}(\mb{g,h}),  
\label{eqn:easy_phi_cocy}
\end{equation}
and
\begin{equation}
	\frac{\eta_a(\mb{g,h})\eta_b(\mb{g,h})}{\eta_c(\mb{g,h})}=\frac{U_\mb{gh}(a,b;c)}{U_\mb{g}(a,b;c)U_\mb{h}( {}^\mb{\ol{g}}a, {}^\mb{\ol{g}}b;{}^\mb{\ol{g}}c)}
	\label{eqn:sliding_consistency}
\end{equation}
It was shown in Ref.~\cite{SET} that $U$ and $\eta$ satisfying Eqs.~\eqref{eqn:FRsym},~\eqref{eqn:easy_phi_cocy} and~\eqref{eqn:sliding_consistency} together with $\rho$ completely determine the symmetry action on anyons.

\twocolumngrid

\section{Defining $U$ and $\eta$ symbols in a chiral CFT}
\label{app:free-boson}
We now explain how to define $U$ and $\eta$ symbols in terms of the symmetry action on local operators in a chiral CFT, generalizing the approach of Ref. \cite{Kawagoe2019}. In fact, our discussions can be adopted with minor modifications to anyons in the (2+1)d bulk. More specifically, the goal is to justify Eqs.~\eqref{eqn:FRsym} and~\eqref{eqn:symloc}, from which Eqs.~\eqref{eqn:easy_phi_cocy} and~\eqref{eqn:sliding_consistency} follow.

For convenience, put the theory on an infinite line $(-\infty, \infty)$.
At the operator level, a chiral primary $a$ is the end point of a topological defect line $V_a$ (in fact, a Verlinde line).  We may also think of $V_a$ as a string operator which commutes with the Hamiltonian except at the end points.  Define $V_a(x)$ to be the string operator running from $-\infty$ to $x$. Define
\begin{equation}
	\ket{a_{x_1}, b_{x_2}, \cdots}=V_a(x_1)V_b(x_2)\cdots \ket{0}.
	\label{}
\end{equation}
It is important that we fix the operators $V_a(x)$ (or the anyonic states $\ket{a_{x_1}, b_{x_2},\dots}$).

To create and manipulate anyonic states, we define a splitting operator $S(a,b;c)$ that transforms the state $\ket{c_{x_1}}$ to $\ket{a_{x_1},b_{x_2}}$:
\begin{equation}
	S(a,b;c)\ket{c_{x_1}}\propto\ket{a_{x_1},b_{x_2}}.
	\label{eqn:splitting}
\end{equation}
We assume that this condition completely fixes $S(a,b;c)$ up to a phase factor. In an MTC, such a splitting operator is abstracted as a vector in the splitting space $V^{ab}_c$. The fact that Eq.~\eqref{eqn:splitting} determines $S$ up to a phase factor means that $V^{ab}_c$ is one-dimensional, or $N_{ab}^c=1$. 

We also need an operator that moves an anyon $a$ from $x_1$ to $x_2$:
\begin{equation}
	M_{x_2,x_1}^a\ket{a_{x_1}}\propto\ket{a_{x_2}}.
	\label{}
\end{equation}

With the splitting and moving operators, one can in principle construct any physical states. 
In Ref.~\cite{Kawagoe2019}, microscopic definitions of $F$ and $R$ symbols using the $S$ and $M$ operators are given.

Let $R_\mb{g}$ be the unitary operator that implements $\mb{g}$ symmetry on the Hilbert space, and $\rho_\mb{g}$ be the corresponding outer automorphism of the chiral algebra.  As $V_a$ can be defined in terms of local operators, we must have
\begin{equation}
	R_\mb{g} V_a(x) R_\mb{g}^{-1}= U_\mb{g}^{a}(x)V_{ {}^\mb{g}a}(x),
	\label{eqn:Uanyon}
\end{equation}
where $U_\mb{g}^a(x)$ is a local unitary operator at $x$.  Eq.~\eqref{eqn:Uanyon} can be replaced by
\begin{equation}
	R_\mb{g}\ket{a_x}= U_\mb{g}^a(x)\ket{{}^\mb{g}a_x}.
	\label{eqn:R-on-anyon}
\end{equation}
The phase of $U_\mb{g}^a(x)$ is ambiguous and may be $x$-dependent as well, but should not depend on other topological charges in the systems as long as they are sufficiently far away.

Under the symmetry $R_\mb{g}$, we must have 
\begin{equation}
	\begin{split}
		R_\mb{g} S(a,b;c)R_\mb{g}^{-1}&=\\
		U_\mb{g}( {}^\mb{g}a, {}^\mb{g}b; {}^\mb{g}c)&U_\mb{g}^a(x_1)U_\mb{g}^b(x_2)S\big({}^\mb{g}a, {}^\mb{g}b; {}^\mb{g}c\big)U_\mb{g}^c(x_1)^\dag.
	\end{split}
	\label{}
\end{equation}

For the moving operator, under the symmetry one finds 
\begin{equation}
	R_\mb{g} M_{x_2,x_1}^aR_\mb{g}^{-1} =  U_\mb{g}^a(x_2)M_{x_2x_1}^{{}^\mb{g}a}U_\mb{g}^a(x_1)^\dag.
	\label{}
\end{equation}
Naively one would include a phase factor for the ``string'' that transports $a$ from $x$ to $x'$. However the ``gauge'' is essentially fixed in Eq.~\eqref{eqn:R-on-anyon}.

Having setting up the microscopic definitions of symmetry action, one can now directly prove Eq.~\eqref{eqn:FRsym} using the definitions of $F$ and $R$ in Ref.~\cite{Kawagoe2019} .

Next, we show that the symmetry localization ansatz Eq.~\eqref{eqn:symloc} indeed holds. Consider the state $\ket{\{a_j\};\{c_j\}}$ with $n$ anyons $a_1,a_2,\dots, a_n$ fusing to the identity, with intermediate fusion channels $c_{1}, c_{2}, \dots, c_{n-1}$.
Suppose $a_j$ is located at $x_j$, with $x_1<x_2<\cdots <x_n$. Using the splitting and moving operators, the state can be constructed explicitly as
\begin{widetext}
\begin{equation}
	\ket{\{a_j\}; \{c_j\}}=S(a_1, a_2;c_1)M_{x_3, x_2}^{a_3}S(c_1, a_3; c_2)\cdots M_{x_{n-1}, x_2}^{a_{n-1}}S(c_{n-2}, a_{n-1}; c_{n-1})M_{x_n, x_2}^{a_n} S(c_{n-1},a_n;0)\ket{0}
	\label{}
\end{equation}
\end{widetext}
Applying $R_\mb{g}$, one finds
\begin{equation}
	R_\mb{g}\ket{\{a_j\}; \{c_j\}}=\prod_{j=1}^n U_\mb{g}^{a_j}(x_j) \rho_\mb{g}\ket{\{a_j\}; \{c_j\}},
	\label{}
\end{equation}
with $\rho_\mb{g}$ given exactly by Eq.~\eqref{eqn:rho_g}, confirming the decomposition in Eq.~\eqref{eqn:symloc}.

In the following we compute $\eta$ and $U$ using these definitions in the chiral $\U_N$ CFT. For simplicity, assume that $N$ is even so the theory is bosonic. The Lagrangian reads
\begin{equation}
	\cal{L}=\frac{N}{4\pi}\partial_t\phi\partial_x\phi - \frac{v}{4\pi}(\partial_x\phi)^2.
	\label{}
\end{equation}
$\phi$ satisfies the following commutation relation:
\begin{equation}
	[\phi(x),\phi(y)]=\frac{i\pi}{N}\mathrm{sgn}(y-x).
	\label{}
\end{equation}
Local operators are generated by $e^{\pm iN\phi}$ and derivatives of $\phi$, which form the $\U_N$ Kac-Moody algebra.

Chiral primaries/anyons are labeled by $l=0,1,\dots, N-1$. We write $[l]$ to mean $l$ mod $N$. Define chiral vertex operators
\begin{equation}
	V_l(x)=e^{il\phi(x)}.
	\label{}
\end{equation}
Following Ref.~\cite{Kawagoe2019}, define the splitting operator
\begin{equation}
	S(l,m)=e^{im\int_{x_1}^{x_2}\di y\,\partial_y\phi}e^{i(l+m-[l+m])\phi(x_1)}.
	\label{}
\end{equation}
Ref.~\cite{Kawagoe2019} used this splitting operator to compute $F$ and $R$ symbols of the $\U_N$ MTC. We work in the same gauge as Ref.~\cite{Kawagoe2019}.

We are concerned with the charge-conjugation symmetry:
\begin{equation}
	C: \phi\rightarrow -\phi,
	\label{}
\end{equation}
which is the only nontrivial outer automorphism of the chiral algebra except for $N=2$ where the chiral algebra becomes SU(2)$_1$. For $N>2$, we have $\mathrm{Aut}(\U_N)=\mathrm{O}(2)$.

 It is easy to find that for $l>0$,
\begin{equation}
	C|l_x\rangle = e^{-iN\phi(x)}|[-l]_x\rangle.
	\label{}
\end{equation}
This is because $[-l]=N-l$. Thus $U_C^l=e^{-iN\phi}$ for $0<l<N$. We then immediately find
\begin{equation}
	\eta_l(C,C)=1.
	\label{}
\end{equation}

Next we compute $U_C(l,m;[l+m])$. Suppose $l,m>0$, we have:
\begin{equation}
	CS(l,m)C^{-1}=e^{-im\phi(x_2)}e^{i([l+m]-l)\phi(x_1)} .
	\label{}
\end{equation}

We also need to compute
\begin{equation}
	S'(l,m)=U_C^l(x_1)U_C^m(x_2)S\big([-l],[-m]\big)U_C^{[l+m]}(x_1)^{-1}.
	\label{}
\end{equation}

Let us first consider $l+m\neq N$:
\begin{equation}
	\begin{split}
	S'(l,m)&=e^{-iN\phi(x_1)}e^{-im\phi(x_2)}e^{i(2N-l-[-l-m])\phi(x_1)}\\
	&=(-1)^me^{-im\phi(x_2)}e^{i(N-l-[-l-m])\phi(x_1)}.
	\end{split}
	\label{}
\end{equation}

We remark that
\begin{equation}
	N-l-[-l-m]
		= [l+m]-l.
	%=\begin{cases}
	%	m & l+m<N\\
	%	m-N & l+m>N
	%\end{cases}
	\label{}
\end{equation}
Thus we obtain
\begin{equation}
	U_C(l,m)=(-1)^m.
	\label{eqn:UC}
\end{equation}

Next we consider $l+m=N$ 
\begin{equation}
	CS(l,N-l)C^{-1}=e^{-i(N-l)\phi(x_2)}e^{-il\phi(x_1)}
	\label{}
\end{equation}
and
\begin{equation}
	\begin{split}
	S'(l,N-l)&=e^{-iN\phi(x_1)}e^{-i(N-l)\phi(x_2)}e^{i(N-l)\phi(x_1)}\\
	&=(-1)^le^{-i(N-l)\phi(x_2)}e^{-il\phi(x_1)} .
	\end{split}
	\label{}
\end{equation}
Therefore $U_C(l,N-l)= (-1)^{l}$, essentially the same answer as in Eq.~\eqref{eqn:UC}. We remark that Ref.~\cite{Kawagoe2019} performed an additional gauge transformation to bring the F symbols to a standard form from the literature (e.g. Ref.~\cite{Moore89b}). With that taken into account, we finally have
\begin{equation}
	U_C(l,m)=(-1)^l, m>0.
	\label{}
\end{equation}
and obviously $U_C(l,0)=1$.

As an example, when $N=2$, the only nontrivial chiral primary is $l=1$, $e^{i\phi}$, corresponding to the semion excitation in the bulk. We find that
\begin{align}
	U_C([1], [1])=-1, && \eta_{[1]}(C,C)=1.
	\label{}
\end{align}
This is gauge-equivalent to
\begin{align}
	U_C([1], [1])=1, && \eta_{[1]}(C,C)=-1.
	\label{}
\end{align}
In other words, the semion has a half $\Z_2$ charge under $C$~\cite{SET}. This is consistent with the orbifold being $\U_8$.

\section{Algebraic description of gapped boundaries}
We review the algebraic theory of gapped boundaries of a two-dimensional topological phase~\cite{kong2014, kitaev2012, lan2015, NeupertPRB2016, Cong2016}. It can be formulated in three different but equivalent ways:

\begin{enumerate}
	\item In the first approach, a gapped boundary corresponds to a Lagrangian algebra of the bulk MTC. Physically the Lagrangian algebra indicates which bulk anyons are condensed on the boundary~\cite{levin2013}.

		We can also study excitations on the boundary. These are of course confined, and since the boundary is one-dimensional it does not make sense to braid such confined excitations. Thus only their fusion properties are of interest, and the (equivalence classes of) boundary excitations form a unitary fusion category (UFC), denoted by $\cal{C}$. The key result in the algebraic theory of gapped boundary is that the bulk is the center of the boundary: the bulk MTC is the Drinfeld center, more commonly known as quantum double to physicists, of the boundary UFC, denoted by $\cal{Z}(\cal{C})$. Physically the Drinfeld center is realized by the generalized string-net models~\cite{Levin05a}.

	\item In the second approach, we use the Drinfeld center as the starting point. A gapped boundary then corresponds to a \emph{module} category over the input UFC $\cal{C}$~\cite{KongRunkel2007, DNO, kitaev2012}. We briefly review the notion of module category below. In a string-net construction, the module category defines string types on the boundary (which can be different from the bulk string types), as well as how bulk strings terminate on the boundary. 

	\item Also starting from a Drinfeld center, a gapped boundary corresponds to a Frobenius algebra $A$ in the UFC $\cal{C}$~\cite{HuJHEP2018}.   
\end{enumerate}

Let us focus on the case relevant for our purpose, namely the input UFC is by itself a MTC $\cal{B}$.  The Drinfeld center is particularly simple: $\cal{Z}(\cal{B})=\cal{B}\boxtimes\ol{\cal{B}}$.  Let us see how to describe this gapped boundary in the formalisms introduced above:
\begin{enumerate}
	\item All ``diagonal'' anyons of the form $(a,a)$ for $a\in \cal{B}$ are condensed on the boundary. So the Lagrangian algebra is $\cal{L}=\sum_{a\in \cal{B}}(a,a)$.
	\item The module category is still isomorphic (set-wise) to $\cal{B}$, with the module action obviously given by the fusion in $\cal{B}$.
	\item The algebra in the UFC is $A=1$, the identity object. 
\end{enumerate}

We can generalize the algebraic descriptions to ``non-diagonal'' condensations as well. For $\varphi\in \text{Aut}(\cal{B})$,
	there is a Lagrangian algebra $\cal{L}=\sum_{a\in \cal{B}}(a,\varphi(a))$.
	However, the module and algebra have to be determined case by case.

	%Let us illustrate with an example. Consider the family of MTCs $\Z_N^{(p)}$, $p$ is an integer (half-integer) for odd (even) $N$. Suppose $\varphi(a)=[-a]_N$.   This is realized by $\sum_{a=0}^{N-1}[a]$ for odd $N$ and $\sum_{a=0}^{N/2-1}[2a]$ for even $N$. Notice that there may be additional algebras. For example, when $N=8$, $[0]+[4]$ is an algebra. One can show that it corresponds to the following Lagrangian algebra
	%\begin{equation}
		%\sum_{a+b \text{ even}} (2a,2b)
	%(0,0)+(0,4)+(4,0)+(4,4)+(2,2)+(2,6)+(6,2)+(6,6).	
		%\label{}
	%\end{equation}
	%Unfolding this gapped boundary gives a \emph{non-invertible} domain wall between two $\Z_8^{(1/2)}$ topological order, and the domain wall can be understood as a thin strip of $\Z_2^{(1/2)}$ phase obtained from $\Z_8^{(1/2)}$ by condensing $[4]$: anyons like $[1]$ in $\Z_8^{(1/2)}$ can not pass through the domain wall.

	\subsection{Definition of condensable algebra}
	
Here we review the algebraic description of gapped boundary as a Lagrangian algebra object in the UMTC, following Ref.~\cite{Cong2017}. The key is to include the local process of annihilating a condensable anyon $a$ on the boundary. Similar to fusion/splitting spaces, we associate a vector space for local operators that annihilate $a$, denoted as $V^a$, with basis vector $\ket{a;\mu}$. We refer to $V^a$ as the boundary condensation space. The dimension of this vector space is exactly the ``multiplicity'' $n_a$ of $a$ in the Lagrangian algebra. Obviously we must have $n_1=1$. 
 
Diagrammatically, the condensation process is represented by an anyon line terminating on a wall representing the boundary. We also attach a label at the termination point which represents the state of the boundary condensation space. When $n_a=1$ it can be suppressed.

An important property of the algebra is the following ``$M$ symbol'':
\begin{equation}
\pspicture[shift=-0.55](-0.2,-0.2)(1.3,1.1)
  \small
  \psset{linewidth=0.9pt,linecolor=black,arrowscale=1.5,arrowinset=0.15}
  \psline{->}(0.2,0)(0.2,0.65)
  \psline(0.2,0)(0.2,0.95)
 \psline{->}(0.8,0)(0.8,0.65)
  \psline(0.8,0)(0.8,0.95)
  \psline(-0.1,0)(1.1,0.0)
  \multido{\n=0.0+0.1}{11}
  {
	  \psline(\n,0)(\n,-0.12)
  }
  \rput[tr]{0}(0.25,-0.16){\scalebox{0.9}{$\mu$}}
 \rput[tr]{0}(0.86,-0.17){\scalebox{0.9}{$\nu$}}
    \rput[bl]{0}(0.9,0.4){$b$}
 \rput[br]{0}(0.1,0.4){$a$}
 %\scriptsize
 % \rput[bl]{0}(0.82,0.37){$\nu$}
  \endpspicture
  =\sum\limits_{c,\lambda}[M^{ab}_c]^{\mu\nu}_\lambda
\pspicture[shift=-0.55](-0.1,-0.2)(1.5,1.2)
  \small
  \psset{linewidth=0.9pt,linecolor=black,arrowscale=1.5,arrowinset=0.15}
  \psline{->}(0.7,0)(0.7,0.45)
  \psline(0.7,0)(0.7,0.55)
  \psline(0.7,0.55) (0.25,1)
  \psline{->}(0.7,0.55)(0.3,0.95)
  \psline(0.7,0.55) (1.15,1)	
  \psline{->}(0.7,0.55)(1.1,0.95)
  \psline(0.1,0)(1.3,0.0)
	\multido{\n=0.2+0.1}{11}
  {
	  \psline(\n,0)(\n,-0.12)
  }
 \rput[tr]{0}(0.76,-0.17){\scalebox{0.9}{$\lambda$}}
  \rput[bl]{0}(0.4,0.2){$c$}
  \rput[br]{0}(1.4,0.8){$b$}
  \rput[bl]{0}(0,0.8){$a$}
 %\scriptsize
 % \rput[bl]{0}(0.82,0.37){$\nu$}
  \endpspicture
.
\end{equation}
Notice one important difference between the $M$ moves and the $F, R$ moves of an anyon model: $F$ and $R$ symbols always represent unitary transformations between different basis states of the same state space. However, here the dimension $n_an_b$ of the left figure does not have be equal to that of the right, which is $\sum_c N_{ab}^c n_c$.  It is shown in Ref.~\cite{Cong2017} that a condensable algebra must satisfy
\begin{equation}
	n_a n_b\leq \sum_c N_{ab}^c n_c.
	\label{}
\end{equation}

Next we impose consistency conditions on the $M$ symbols. We can apply $M$ moves to three anyon lines terminating $a,b,c$ on the boundary, but in different orders, which leads to a variation of the pentagon equation: 
\begin{equation}
	\sum_{e,\sigma}[M^{ab}_e]^{\mu\nu}_\sigma [M^{ec}_d]^{\sigma\lambda}_{\delta}[F^{abc}_d]_{ef}=\sum_{\psi}[M^{af}_{d}]^{\mu\psi}_\delta [M^{bc}_f]^{\nu\lambda}_\psi 
	\label{}
\end{equation}
In writing down this equation we assume that the anyon model has no fusion multiplicities, but the generalization is obvious.

The $M$ symbols also have gauge degrees of freedom, originating from the basis transformation of the boundary condensation space $V^a$: $\widetilde{\ket{a;\mu}}=\Gamma^a_{\mu\nu}\ket{a;\nu}$, where $\Gamma^a_{\mu\nu}$ is a unitary transformation. The $M$ symbol becomes
\begin{equation}
	[\tilde{M}^{ab}_c]^{\mu\nu}_\lambda=\sum_{\mu',\nu',\lambda'}\Gamma^{a}_{\mu\mu'}\Gamma^b_{\nu\nu'}[M^{ab}_c]^{\mu'\nu'}_{\lambda'}[\Gamma^c]_{\lambda'\lambda}^{-1}.
	\label{}
\end{equation}
$M$ symbols are affected by the gauge transformation of bulk fusion space as well.

It is convenient to fix the gauge for the following symbols:
\begin{equation}
	[M^{1a}_a]^{\mu}_\nu=[M^{a1}_a]^{\mu}_\nu=\delta_{\mu\nu}.
	\label{}
\end{equation}

Braiding puts further constraints on the $M$ symbols. Since the anyons condense on the boundary, it should not matter in which order the anyon lines terminate on the boundary. Diagrammatically, we have 
\begin{equation}
  \pspicture[shift=-0.6](-1.0,-0.05)(1.8,2.0)
  \small
  \psset{linewidth=0.9pt,linecolor=black,arrowscale=1.5,arrowinset=0.15}
  \psbezier(1,0) (1,0.75) (0,0.45) (0,1.1)
	\psbezier[border=2.2pt](0,0) (0,0.75) (1,0.45) (1,1.1)
\psline(1,1.1) (1,1.6)
	\psline{->}(1,1.1) (1,1.4)
\psline(0,1.1) (0,1.6)
	\psline{->}(0,1.1) (0,1.4)
  \psline(-0.5,0)(1.5,0.0)
	\multido{\n=-0.4+0.1}{19}
  {
	  \psline(\n,0)(\n,-0.12)
  }
\rput[tr]{0}(-0.1, 1.6){$a$}
\rput[tl]{0}(1.1, 1.6){$b$}
  \rput[tr]{0}(0.06,-0.17){\scalebox{0.9}{$\nu$}}
 \rput[tr]{0}(1.05,-0.16){\scalebox{0.9}{$\mu$}}
	%\psbezier[linewidth=0.9pt, border=2pt](1.5,2.2) (1.0,1.2) (-0.1,2.3) (-0.5,0)
\endpspicture
=
  \pspicture[shift=-0.6](-0.8,-0.05)(1.8,2.0)
  \small
	\psset{linewidth=0.9pt,linecolor=black,arrowscale=1.5,arrowinset=0.15}
	\psline(1,0) (1,1.6)
	\psline{->}(1,0) (1,1)
	\psline(0,0) (0,1.6)
	\psline{->}(0,0) (0,1)
  \psline(-0.5,0)(1.5,0.0)
	\multido{\n=-0.4+0.1}{19}
  {
	  \psline(\n,0)(\n,-0.12)
  }
\rput[tr]{0}(-0.1, 1.6){$a$}
\rput[tl]{0}(1.1, 1.6){$b$}
  \rput[tr]{0}(0.05,-0.16){\scalebox{0.9}{$\mu$}}
 \rput[tr]{0}(1.06,-0.17){\scalebox{0.9}{$\nu$}}
	%\psbezier[linewidth=0.9pt, border=2pt](1.5,2.2) (1.0,1.2) (-0.1,2.3) (-0.5,0)
\endpspicture,
\end{equation}
which leads to the following:
\begin{equation}
	[M^{ba}_c]^{\nu\mu}_\lambda R^{ab}_c=[M^{ab}_c]^{\mu\nu}_\lambda.
	\label{eqn:Mbraiding}
\end{equation}
There is a similar condition for the inverse braiding.

It was shown in Ref.~\cite{Cong2017} that these conditions are equivalent to the mathematical definition of a commutative, connected and separable Frobenius algebra $\mathcal{A}=\bigoplus_a n_a a$ in a braided tensor category,  with the algebra morphism $\mathcal{A}\times \mathcal{A}\rightarrow \mathcal{A}$ precisely given by the $M$ symbol. 

%Let us now consider the example of doubled topological phase $\cal{B}\boxtimes\ol{\cal{B}}$, with the Lagrangian algebra $\sum_{a\in \cal{B}}(a,a)$. We write $M^{(a,a),(b,b)}_{(c,c)}\equiv M^{ab}_c$ for $N_{ab}^c>0$.
%The F and R symbols for condensable anyons are all $1$, so we have two conditions
%\begin{equation}
%	\sum_{e}M^{ab}_e M^{ec}_d= M^{bc}_fM^{af}_d, M^{ab}_c=M^{ba}_c.
%	\label{}
%\end{equation}
%Thus an obvious solution is $M^{ab}_c=\frac{\sqrt{d_c}{d_ad_b}}N^{ab}^c$. We believe this is the unique solution up to gauge transformations.

\subsection{Symmetry-preserving condensation}
\label{sec:charged-condensate}
We now give a precise definition of anyon condensation that preserves the global symmetry~\cite{Bischoff2019}. Denote by $\cal{L}$ the Lagrangian algebra (the discussion applies to a general commutative algebra as well). In the following $a,b, c, \dots$ denote anyons in the condensate, unless otherwise specified. We assume $n_a=1$ whenever $a$ belongs to the condensate, so we omit the index for the boundary condensation space.

We draw diagrams where $\mb{g}$ defect lines terminate on the boundary. Strictly speaking, the defect line should continue into an SPT phase (unless the 't Hooft anomaly described in Sec.~\ref{sec:SETboundary} vanishes) and in principle should be described by a theory of condensation in a $G$-crossed braided category. However, we leave this for future work and proceed more heuristically.  Since the boundary is fully gapped and symmetric, we can posit that for each $\mb{g}$ there exists at least one $\mb{g}$-defect that can be absorbed  without creating any additional excitations on the boundary. Thus the $M$ move is also defined for these defects:
\begin{equation}
\pspicture[shift=-0.55](-0.2,-0.2)(1.3,1.1)
  \small
  \psset{linewidth=0.9pt,linecolor=black,arrowscale=1.5,arrowinset=0.15}
  \psline{->}(0.2,0)(0.2,0.65)
  \psline(0.2,0)(0.2,0.95)
 \psline{->}(0.8,0)(0.8,0.65)
  \psline(0.8,0)(0.8,0.95)
  \psline(-0.1,0)(1.1,0.0)
  \multido{\n=0.0+0.1}{11}
  {
	  \psline(\n,0)(\n,-0.12)
  }
  \rput[bl]{0}(0.92,0.4){$y_\mb{h}$}
	\rput[br]{0}(0.08,0.4){$x_\mb{g}$}
 %\scriptsize
 % \rput[bl]{0}(0.82,0.37){$\nu$}
  \endpspicture
  =\sum\limits_{z_\mb{gh}}[M^{x_\mb{g}y_\mb{h}}_{z_\mb{gh}}]
\pspicture[shift=-0.55](-0.1,-0.2)(1.5,1.2)
  \small
  \psset{linewidth=0.9pt,linecolor=black,arrowscale=1.5,arrowinset=0.15}
  \psline{->}(0.7,0)(0.7,0.45)
  \psline(0.7,0)(0.7,0.55)
  \psline(0.7,0.55) (0.25,1)
  \psline{->}(0.7,0.55)(0.3,0.95)
  \psline(0.7,0.55) (1.15,1)	
  \psline{->}(0.7,0.55)(1.1,0.95)
  \psline(0.1,0)(1.3,0.0)
	\multido{\n=0.2+0.1}{11}
  {
	  \psline(\n,0)(\n,-0.12)
  }
  \rput[bl]{0}(0.8,0.2){$z_\mb{gh}$}
  \rput[br]{0}(1.55,0.9){$y_\mb{h}$}
  \rput[bl]{0}(-0.15,0.9){$x_\mb{g}$}
 %\scriptsize
 % \rput[bl]{0}(0.82,0.37){$\nu$}
  \endpspicture
.
\end{equation}

We introduce the following move:
\begin{equation}
\pspicture[shift=-0.55](-0.2,-0.2)(1.7,1.5)
  \small
  \psset{linewidth=0.9pt,linecolor=black,arrowscale=1.5,arrowinset=0.15}
  \psline{->}(0.5,0)(0.5,1)
  \psline(0.5,0)(0.5,1.4)
  \psbezier[border=2.2pt](0,0) (0,0.65) (1,0.4) (1,1.0)
\psline(1,1.0) (1,1.4)
	\psline{->}(1,1.0) (1,1.2)
  %\psline[border=2pt](0,0) (1.1,1.1)
  %\psline{->}(0,0) (0.9,0.9)
  \psline(-0.2,0)(1.5,0.0)
  \multido{\n=-0.1+0.1}{16}
  {
	  \psline(\n,0)(\n,-0.12)
  }
 \rput[br]{0}(0.37,0.8){$a$}
\rput[bl]{0}(1.15,0.8){$x_\mb{g}$}
 %\scriptsize
 % \rput[bl]{0}(0.82,0.37){$\nu$}
  \endpspicture
  =\chi_a(\mb{g})
\pspicture[shift=-0.55](-0.3,-0.2)(2.0,1.5)
  \small
  \psset{linewidth=0.9pt,linecolor=black,arrowscale=1.5,arrowinset=0.15}
  \psline{->}(0.5,0)(0.5,1)
  \psline(0.5,0)(0.5,1.4)
  \psline(1,0) (1,1.4)
\psline{->}(1,0) (1,1)
  \psline(-0.2,0)(1.5,0.0)
  \multido{\n=-0.1+0.1}{16}
  {
	  \psline(\n,0)(\n,-0.12)
  }
 \rput[br]{0}(0.37,0.8){$a$}
 \rput[bl]{0}(1.15,0.8){$x_\mb{g}$}
 %\scriptsize
 % \rput[bl]{0}(0.82,0.37){$\nu$}
  \endpspicture
.
\end{equation}
Here $\chi_a(\mb{g})$ is a phase factor. Physically, $\chi_a(\mb{g})$ encodes the $\mb{g}$ action on the condensed anyon. When there is more than one condensation channel, $\chi$ should be replaced by a unitary transformation acting on the condensation space. 

If we slide a vertex which splits a $\mb{gh}$-defect to $\mb{g}$- and $\mb{h}$-defects over a boundary vertex, we find
\begin{equation}
	\eta_{a}(\mb{g,h})=\frac{\chi_a(\mb{gh})}{\chi_{ {}^{\bar{\mb{g}}}a}(\mb{h})\chi_a(\mb{g})}.
	\label{eqn:defect-bdr1}
\end{equation}

We can also consider fusion of condensable anyons on a boundary. For $a,b,c$ in the condensate, sliding a $\mb{g}$ line over the diagrammatic equation that defines $M$ symbol, one finds
\begin{equation}
	M^{{}^{\bar{\mb{g}}}a, {}^{\bar{\mb{g}}}b}_{{}^{\bar{\mb{g}}}c}U_\mb{g}(a,b;c)= M^{ab}_c \frac{\chi_a(\mb{g})\chi_b(\mb{g})}{\chi_c(\mb{g})}.
	\label{eqn:defect-bdr2}
\end{equation}

We believe that these two conditions Eqs.~\eqref{eqn:defect-bdr1} and~\eqref{eqn:defect-bdr2} are sufficient and necessary for the condensation to preserve symmetry.  Mathematically, $\chi_a(\mb{g})$ defines an algebra isomorphism for each $\mb{g}$. The consistency conditions guarantee that one has a $G$-equivariant algebra structure on $\cal{L}$~\cite{Bischoff2019}.

We first consider the case when $\rho\equiv \mathds{1}$, then the equations simplify:
\begin{equation}
	\begin{gathered}
	\eta_a(\mb{g,h})=\frac{\chi_a(\mb{gh})}{\chi_{ a}(\mb{h})\chi_a(\mb{g})},\\
	\chi_a(\mb{g})\chi_b(\mb{g})=\chi_c(\mb{g}), \text{ for }N_{ab}^c>0.
	\end{gathered}
	\label{}
\end{equation}
It then follows that one can write $\chi_{a}(\mb{g})=M_{a,\cohosub{t}(\mb{g})}$ for $\coho{t}(\mb{g})\in \cal{A}_\cal{C}$. $\coho{t}$ is not uniquely determined, as one can freely change $\coho{t}$ by an Abelian anyon in the condensate without affecting $\chi_a$. Therefore, $\coho{t}$ belongs to $\cal{A}'=\cal{A}_\cal{C}/\cal{A}_\cal{L}$ where $\cal{A}_\cal{L}$ is the group of condensed Abelian anyons.  Since $\eta_a(\mb{g,h})=M_{a,\cohosub{w}(\mb{g,h})}$ with $\coho{w}(\mb{g,h})\in \cal{A}$, for our purpose we can project $\coho{w}$ to $\cal{A}'$ as well, and the projection will be denoted by $\coho{w}'$. The condition is that $\coho{w}'$ is a trivial 2-cocycle in $\H^2[G, \cal{A}']$, in agreement with the result of Ref.~\cite{Bischoff2019}. 

Back to the general case, notice that these two equations do not fix $\chi$'s uniquely: there is a freedom to change $\chi$ by  $\chi_a(\mb{g})\rightarrow \chi_a(\mb{g})\phi_a(\mb{g})$ where
\begin{align}
	\phi_a(\mb{g})\phi_b(\mb{g})=\phi_c(\mb{g}), && \phi_{ {}^{\bar{\mb{g}}}a}(\mb{h})\phi_a(\mb{g})=\phi_a(\mb{gh}).
	\label{eqn:phi-bdr}
\end{align}
Again we can write $\phi_x(\mb{g})=M_{x,\cohosub{t}(\mb{g})}$, and the two equations reduce to $\coho{t}(\mb{g}) {}^\mb{g}\coho{t}(\mb{h})=\coho{t}(\mb{gh})$. Notice that $\coho{t}(\mb{g})$ is defined up to Abelian anyons in the Lagrangian algebra.
Therefore, the solutions are classified by $\H^1_\rho[G, \cal{A}']$.

For a doubled SET phase $\cal{C}=\cal{B}\boxtimes\ol{\cal{B}}$, and $\cal{L}=\sum_{a\in \cal{B}}(a,a)$, we have $\cal{A}_\cal{C}=\cal{A}\times\cal{A}$ where $\cal{A}$ is the group of Abelian anyons in $\cal{B}$, and $\cal{A}_{\cal{L}}=\cal{A}$. Thus $\cal{A}'=\cal{A}\times\cal{A}/\cal{A}=\cal{A}$, as expected.

\section{Symmetry-enriched string-net models}
We review the generalized string-net models, which can realize all (non-anomalous) symmetry-enriched quantum double topological orders~\cite{HeinrichPRB2016,ChengPRB2017}. The input is a $G$-graded fusion category $\cal{C}_G$:
\begin{equation}
	\cal{C}_G=\bigoplus_{\mb{g}\in G} \cal{C}_\mb{g}.
	\label{}
\end{equation}
Denote the simple objects by $a_\mb{g}, b_\mb{h}, \dots$ etc. The $G$-grading implies that
\begin{equation}
	a_\mb{g}\times b_\mb{h}=\sum_{c_\mb{gh}\in \cal{C}_\mb{gh}}N_{a_\mb{g}b_\mb{h}}^{c_\mb{gh}}c_\mb{gh}.
	\label{}
\end{equation}

Let us now specify the Hilbert space of the model. Each edge of the graph is associated with a Hilbert space whose orthonormal basis is labeled by simple objects in $\cal{C}_G$. To account for the symmetry, we add a spin degree of freedom in the center of each plaquette, whose basis states $\ket{\mb{g}}$ are labeled by the elements $\mb{g}$ of the symmetry group $G$. For each edge we can then associate a group element $\bar{\mb g}_1\mb{g}_2$ (here $\bar{\mb g}$ denotes the inverse of group element $\mb g$). The labels on the edge must belong to the $\cal{C}_{\bar{\mb g}_1\mb{g}_2}$ sector, otherwise they induce an energy penalty:

\begin{equation}
 \begin{tikzpicture}[baseline={($ (current bounding box) - (0,3.5pt) $)}]
	 \draw (0, 0) node [above] {$a_{\ol{\mb{g}}_1\mb{g}_2}$} -- (0, -1.0) [anyon];
		\node  at (-0.3, -0.4) {\scalebox{0.8}{$\mb{g}_1$}};
		\node  at (0.35, -0.4) {\scalebox{0.8}{$\mb{g}_2$}};
		\draw[dashed, thick] (-0.5, -0.75) -- (0.5, -0.75) [dw];
\end{tikzpicture} .
\end{equation}

Three strings meet at a vertex. Whether three string types $a$, $b$ and $c$ are allowed to meet at a vertex or not is determined by the fusion rule $N^{ab}_c$, which is a non-negative integer. If $N^{ab}_c>0$, $a,b$ and $c$ can meet at a vertex without costing an energy penalty:
\begin{equation}
	\begin{tikzpicture}[baseline={($ (current bounding box) - (0,0pt) $)}]
		\draw (0, .5) node [above] {$a$}--(.5, 0) [anyon];
		\draw (1, .5) node [above] {$b$}--(.5, 0) [anyon];
		\draw (.5, 0)--(.5, -.5) node [below] {$c$} [anyon];
	\end{tikzpicture} .
	\label{}
\end{equation}
When $N^{ab}_c>1$, one has to include additional local degrees of freedom at each vertex. We assume $N^{ab}_c$ only takes values in $\{0,1\}$ to simplify the discussions. 

The ground-state wave function is a superposition of string-net states (i.e., string states on the lattice that satisfy the branching rules). A defining feature of the string-net wave function is that the amplitudes for different string-net states satisfy a set of local relations. Most importantly:
	\begin{equation}
    \label{eq:fmove}
    \Psi\left(\begin{tikzpicture}[baseline={($ (current bounding box) - (0,0pt) $)}, scale=0.8]
		\draw (0, 0) node [above] {$\ag ag$} --(.3, -.5) [anyon];
		\draw (.6, 0) node [above] {$\ag bh$} --(.3, -.5) [anyon];
		\draw (.3, -.5)--(.6, -1) [anyon]
		      node [midway, left] {$\ag e{gh}$};
		\draw (1.2, 0) node [above] {$\ag ck$} --(.6, -1) [anyon];
		\draw (.6, -1)--(.9, -1.5) [anyon] node [below] {$\ag d{ghk}$};
		\end{tikzpicture}\right)
		=\sum_{f_\mb{hk}\in \mathcal{C}_\mb{hk}}
		[F^{\ag ag\ag bh\ag ck}_{\ag d{ghk}}]_{\ag e{gh}\ag f{hk}}
    \Psi\left(\begin{tikzpicture}[baseline={($ (current bounding box) - (0,0pt) $)}, scale=0.8]
		\draw (0, 0) node [above] {$\ag ag$} --(.6, -1) [anyon];
		\draw (.6, 0) node [above] {$\ag bh$} --(.9, -.5) [anyon];
		\draw (.9, -.5)--(.6, -1) [anyon]
		      node [midway, right] {$\ag f{hk}$};
		\draw (1.2, 0) node [above] {$\ag ck$} --(.9, -.5) [anyon];
		\draw (.6, -1)--(.9, -1.5) [anyon] node [below] {$\ag d{ghk}$};
		\end{tikzpicture}\right).
  \end{equation}
Here $F$ is the F symbol of the input UFC. We have left the group labels in plaquettes implicit in the picture, as the move only depends on the group labels on edges.

Following Refs.~\cite{HeinrichPRB2016,ChengPRB2017} one can write down a commuting projector Hamiltonian for the ground-state wavefunction. There are both edge and vertex projectors to enforce the $G$-grading and fusion rules, accompanied by terms that fluctuate both the spin on the plaquette together with the surrounding string-net configuration. The resulting topological order is the quantum double of the identity component $\cal{C}_\mb{1}$, and the full $G$-crossed braided fusion category describing the emergent SET order can be extracted from the ground-state wavefunction following Ref.~\cite{WilliamsonSET}. 

\section{String-net models with gapped boundary}
In this appendix we consider generalized string-net models with fully gapped boundary. We assume the boundary is ``smooth'', i.e. no dangling edges, and discuss the case without $G$ grading first. For now we denote the input UFC just by $\cal{C}$. It is important to realize that boundary edges can have a different set of labels from those in the bulk, as long as one consistently defines how the bulk strings terminate on the boundary.   More precisely, one has to specify the following fusion rules between bulk and boundary edges: 
\begin{equation}
	a\times {\alpha}=\sum_{\beta\in \cal{M}}\tilde{N}_{a,\alpha}^\beta \beta.
	\label{eqn:moduleaction}
\end{equation}
Here Greek letters $\alpha, \beta, \dots$ denote labels allowed on boundary edges. They are said to form a (left) module category $\cal{M}$ over the UFC $\cal{C}$~\cite{kitaev2012}, with the module action given by Eq.~\eqref{eqn:moduleaction}. 

\begin{figure}
\centering
\begin{tikzpicture}[scale=0.25,>=stealth]
\def\shift{14};
    %\draw[help lines] (0,0) grid (26,8);
  \draw[anyon] (0,8) node[above]{$a$} -- (2,6) ;
    \draw[anyon]  (2,6) -- (4,4);
    \draw[anyon]  (2,8) node[above]{$b$}-- (2,6);
    \draw[very thick, dotted ] (4,4) -- (4,2) node[below]{$\beta$} ;
\node[draw=none,fill=none] at (3.2,5.7) {$c$}; 
    \draw[very thick, dotted](4,8) node[above]{$\alpha$}-- (4,4) ;
	
	\node[draw=none,fill=none] at (10,5) {$=\displaystyle\sum_{\gamma\in\cal{M}}[m^{ab\alpha}_{\beta}]_{c\gamma}$}; 
    \draw[anyon] (\shift+0,8) node[above]{$a$}-- (\shift+4,4) ;
    \draw[anyon]  (\shift+2,8) node[above]{$b$}-- (\shift+4,6) ;
    \draw[very thick, dotted,-] (\shift+4,4) -- (\shift+4,2)node[below]{$\beta$};
	\node[draw=none,fill=none] at (\shift+4.7,5) {$\gamma$}; 
    \draw[very thick, dotted, -](\shift+4,8) node[above]{$\alpha$}-- (\shift+4,4);
\end{tikzpicture}
\caption{The $m$ move in a left $\mathcal{C}$-module $\cal{M}$. The dashed lines represent boundary edges.}
\label{fig:mmove}
\end{figure}
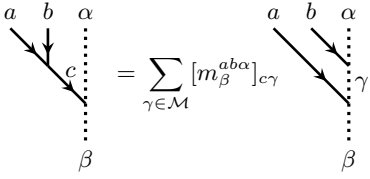

In order to consistently define the wavefunction, the F move needs to be extended to include boundary edges. A boundary F move is depicted in Fig.~\ref{fig:mmove}. We denote the transformation by $m$.  The ``pentagon'' equation now reads
\begin{equation}
	\sum_f (F^{abc}_d)_{ef} (m^{af\alpha}_\beta)_{d\gamma} (m^{bc\alpha}_\gamma)_{f\delta}=(m^{ec\alpha}_\beta)_{d\lambda}(m^{ab\lambda}_\beta)_{e\gamma}.
	\label{}
\end{equation}

The $m$ symbols are defined up to gauge transformations:
\begin{equation}
	(m^{ab\alpha}_\beta)_{c\gamma}\rightarrow \frac{v^{b\alpha}_\gamma v^{a\gamma}_\beta}{v^{c\alpha}_\beta}(m^{ab\alpha}_\beta)_{c\gamma}.
	\label{}
\end{equation}
 We also assume that the $m$ symbols are normalized:
\begin{align}
	(m^{1b\alpha}_\beta)_{b,\beta}=1, && (m^{a1\alpha}_\beta)_{a,\alpha}=1.
	\label{}
\end{align}
These definitions are completely parallel to the usual $F$ symbols.

Technically, what we have defined is a left module category. Right module categories can be defined analogously.

For example, we may take the labels of $\cal{M}$ to be the same as those of $\cal{C}$, and the module action to be given by fusion rules in $\cal{C}$, $m$ is then equal to $F$ and the pentagon equation is automatically satisfied. This module defines the canonical ``smooth'' boundary for a string-net model.

Now we turn to the symmetry-enriched models. Mathematically, this question was addressed in Ref.~\cite{Meir}. We will however proceed with a more heuristic physical construction. An immediate problem to resolve is how to make the $G$-grading compatible with a gapped boundary. Naively, one may imagine adding additional $G$ spins directly outside the boundary, but this construction of this sort either lead to extensively degenerate boundary states or spontaneous symmetry breaking~\cite{LevinGu}. Instead, we do not assign any $G$-graded structure to the module category directly. Thus the module action reads
\begin{equation}
	a_\mb{g}\times \alpha=\sum_{\beta\in \cal{M}}\tilde{N}_{a_\mb{g},\alpha}^\beta \beta.
	\label{}
\end{equation}
This means that $\cal{M}$ is in fact a module category over $\cal{C}_0$, the identity component, which is then ``upgraded'' to a module over the whole $\cal{C}_G$. Namely, both the module action and the $m$ moves must be extended to the whole $G$-graded category. If this is not possible, the gapped boundary to vacuum labeled by $\cal{M}$ breaks the symmetry. In some cases the symmetry can be restored by considering $\cal{M}$ as a gapped boundary to an appropriate SPT phase that matches the anomaly introduced in Sec.~\ref{sec:SETboundary}. Physically, one may picture the strings with nonzero $\mb{g}$ as representing domain wall configurations, and a symmetric gapped boundary should allow domain walls and certain symmetry defects to ``condense''. 

While we focus on the construction based on module category, we remark that the dual algebra description allows one to calculate these symmetry defects that condense at a symmetric boundary in a straightforward manner. This phenomena was noted in an example in Ref.~\cite{Kesselring2018}. We plan to expound on this topic in future work. 

\subsection{Gapped boundaries in symmetry-enriched doubled topological phases}
\label{sec:set-model}

We consider an MTC $\cal{B}$, and its quantum double $\cal{Z}(\cal{B})=\cal{B}\boxtimes \ol{\cal{B}}$. We assume that the symmetry action is ``diagonal'', in other words the SET order in $\ol{\cal{B}}$ is the conjugate of that in $\cal{B}$. It is then sufficient to specify the symmetry-action data $\rho, U, \eta$ in $\cal{B}$.
Our goal is to compute $\omega$ in terms of  $\rho, U, \eta$ and the charges carried by the condensed anyons, parametrized by $\coho{v}$. 

The first task is to construct a suitable $G$-graded extension $\cal{B}_G$ of the fusion category $\cal{B}$. 
 We postulate that $\cal{B}_\mb{g}$ are all ``copies'' of $\cal{B}$. In other words, we can write
\begin{equation}
	\cal{B}_\mb{g}=\{a_\mb{g}|a\in \cal{B}\}.
	\label{}
\end{equation}
We consider the following fusion rules
\begin{equation}
	a_\mb{g}\times b_\mb{h} =\sum_{c\in\cal{B}}N_{a,{}^\mb{g}b}^c \,c_\mb{gh}.
	\label{eqn:setsn-fusion}
\end{equation}
Such a fusion rule is motivated by the semi-direct product for group extensions.  We further postulate the following F symbol:
\begin{equation}
	\begin{split}
		[F^{a_\mb{g}, b_\mb{h},c_\mb{k}}_{d_\mb{ghk}}&]_{e_\mb{gh}, f_\mb{hk}} =[F^{a, {}^\mb{g}b, {}^\mb{gh}c}_d]_{e, {}^\mb{g}f}\\
	&U_\mb{g}^{-1}( {}^\mb{g}b, {}^\mb{gh}c;  {}^\mb{g}f)\eta_{ {}^\mb{gh}c}(\mb{g,h}).
	\end{split}
	\label{eqn:setsn-F}
\end{equation}
Using Eqs.~\eqref{eqn:FRsym},~\eqref{eqn:easy_phi_cocy} and~\eqref{eqn:sliding_consistency}, one can prove that the F symbols indeed satisfy the pentagon equation. It remains to show that the graded extension specified by Eqs.~\eqref{eqn:setsn-fusion} and~\eqref{eqn:setsn-F} indeed gives the doubled SET phase, which we postpone for now.

Let us study module categories over $\cal{B}_G$. The fact that each $\cal{B}_\mb{g}$ is a ``copy'' of $\cal{B}$, at least where the fusion rules are concerned, motivates the following form of module action:
\begin{equation}
	\begin{split}
		x_\mb{g}\times \tilde{a} &= \sum_{b\in \cal{B}_\mb{0}}N_{x, {}^\mb{g}a}^{b}\,\tilde{b}.
	\end{split}
	\label{}
\end{equation}
Here we use $\tilde{}$ to denote boundary labels (since they are the same \emph{set} as the bulk ones). The module $m$ symbol is just the bulk $F$ symbol:
\begin{equation}
	[m^{a_\mb{g}b_\mb{h}\tilde{c}}_{\tilde{d}}]_{e_\mb{gh}, \tilde{f}}=[F^{a_\mb{g}b_\mb{h}c_\mb{0}}_{d_\mb{gh}}]_{e_\mb{gh}, f_\mb{h}}.
	\label{}
\end{equation}
It is easy to check that the $m$ symbols given above do satisfy the boundary pentagon equation. Therefore we have at least constructed one symmetry-preserving module category over $\cal{B}_G$, which shows the symmetry-enriched string-net model does have a symmetric gapped boundary to vacuum.

Next we observe that the module action can be ``twisted'' in the following way:
\begin{equation}
	\begin{split}
		x_\mb{g}\times \tilde{a} &= \sum_{b\in \cal{B}_\mb{0}}N_{x,{}^\mb{g}a}^{b\times v(\mb{g})}\,\tilde{b}.
	\end{split}
	\label{eqn:twisted-module-action}
\end{equation}
Here $v(\mb{g})\in \cal{A}$. 
Associativity of the module action requires 
\begin{equation}
v(\mb{g})\times {}^\mb{g}v(\mb{h})=v(\mb{gh}).
	\label{}
\end{equation}
Thus $v$ defines a twisted homomorphism from $G$ to $\cal{A}$.

To check whether the twisted module action actually defines a module category, we need to solve for the $m$ symbols.  Consider the boundary pentagon equation for $[{{v}(\mb{g})}]_\mb{g}, [{{v}(\mb{h})}]_\mb{h}, [{{v}(\mb{k})}]_\mb{k}, \tilde{0}$. Note that $[v(\mb{g})]_\mb{g}\times\tilde{0}=\tilde{0}$. The boundary pentagon equation reads
\begin{equation}
	F^{[v(\mb{g})]_\mb{g},[v(\mb{h})]_\mb{h},[v(\mb{k})]_\mb{k}}=
	\frac{ m^{{v}(\mb{gh})_\mb{gh},{v}(\mb{k})_\mb{k},\tilde{0}}_{\tilde{0}} m^{{v}(\mb{g})_\mb{g},{v}(\mb{h})_\mb{h},\tilde{0}}_{\tilde{0}} }
	{m^{{v}(\mb{g})_\mb{g},{v}(\mb{hk})_\mb{hk},\tilde{0}}_{\tilde{0}} m^{{v}(\mb{h})_\mb{h},{v}(\mb{k})_\mb{k},\tilde{0}}_{\tilde{0}}}.
	\label{}
\end{equation}
It means that $F^{[{v}(\mb{g})]_\mb{g},[{v}(\mb{h})]_\mb{h},[{v}(\mb{k})]_\mb{k}}$ must be a cohomologically trivial $3$-cocycle in $\H^3[G, \U]$. Thus we interpret $F^{[{v}(\mb{g})]_\mb{g},[{v}(\mb{h})]_\mb{h},[{v}(\mb{k})]_\mb{k}}$ as the 3-cocycle that defines the SPT phase: 
\begin{multline}
	F^{[{v}(\mb{g})]_\mb{g},[{v}(\mb{h})]_\mb{h},[{v}(\mb{k})]_\mb{k}}=F^{v(\mb{g}), {}^\mb{g}v(\mb{h}), {}^\mb{gh}v(\mb{k})}\\
	\cdot U_\mb{g}^{-1}\big({}^\mb{g}v(\mb{h}), {}^\mb{gh}v(\mb{k})\big) \eta_{ {}^\mb{gh}v(\mb{k})}(\mb{g,h}).
	\label{eqn:anomaly4}
\end{multline}

Physically, we believe that $v(\mb{g})$ describe the charge assignment on condensed anyons. Intuitively, the twisted module action Eq. \eqref{eqn:twisted-module-action} means that when $\mb{g}$ domain walls end on the boundary, additional $v(\mb{g})$ anyon labels are attached. This agrees with the physical picture of symmetry defects dressed by $\coho{v}(\mb{g})$ anyons in order to ``pass through'' the gapped boundary described in Sec. \ref{sec:SETboundary}. In addition, for the special case where $\rho\equiv \mathds{1}$ and $U, \eta$ all equal to $1$, the above formula Eq. \eqref{eqn:anomaly4} agrees with Eq. \eqref{eqn:anomaly1}. Therefore we identify $v(\mb{g})$ with $\coho{v}(\mb{g})$, and the Eq. \eqref{eqn:anomaly4} is the formula for relative 't Hooft anomaly.

Now we study the SET order that emerges from the generalized string-net construction defined by Eqs.~\eqref{eqn:setsn-fusion} and~\eqref{eqn:setsn-F}. It can be worked out following the procedure described in Ref.~\cite{WilliamsonSET}. We now sketch the derivation for $\rho\equiv \mathds{1}$. The idea is to study the ground state space of the symmetry-enriched string-net model on a cylinder, or equivalently an annulus, with open boundary conditions. Since the string-net model produces a fixed-point wavefunction, it is enough to consider a basis of ``minimal'' configurations:
\begin{align*}
\begin{array}{c}
\begin{tikzpicture}[scale=.1]
\def\dx{-1};
\def\ddx{1.5};
\def\dy{.5};
\def\p{5.5};
\filldraw[UFCBackground](0,0)--(20,0)--(20,20)--(0,20)--(0,0);
\draw[anyon](13,9)--(20,9) node[pos=0.5,above] {$a_\mb{g}$};
\draw[anyon](7,11)--(13,9) node[pos=0.25,below] {$b_\mb{g}$};
\draw[anyon](0,11)--(7,11) node[pos=0.55,above] {$a_\mb{g}$};
\draw[anyon](10,0)--(13,9)  node[pos=0.3,right] {$c_\mb{1}$};
\draw[anyon](7,11)--(10,20) node[pos=0.55,right] {$c_\mb{1}$};
\draw[draw=black!29,dashed](0,0)--(20,0) (20,20)--(0,20);
\filldraw[draw=black!29,fill=black!29](\p,0)--(\p+\dx,-\dy)--(\p+\dx,\dy)--(\p,0);
\filldraw[draw=black!29,fill=black!29](\p,20)--(\p+\dx,20-\dy)--(\p+\dx,20+\dy)--(\p,20);
\filldraw[draw=black!29,fill=black!29](\p+\ddx,0)--(\p+\dx+\ddx,-\dy)--(\p+\dx+\ddx,\dy)--(\p+\ddx,0);
\filldraw[draw=black!29,fill=black!29](\p+\ddx,20)--(\p+\dx+\ddx,20-\dy)--(\p+\dx+\ddx,20+\dy)--(\p+\ddx,20);
	\end{tikzpicture}
\end{array}, 
\end{align*}
where the top and bottom edges are identified. These figures represent minimal string-net states with a $\mb{g}$ defect line through the cylinder.
Under concatenation these cylinders form the so-called defect tube (dube) algebra. To identify the superselection sectors of the emergent symmetry-enriched topological order, we compute the irreducible central idempotents of the dube algebra. The states formed by these idempotents correspond to minimally entangled states with $\mb{g}$ flux and also definite topological charge through the cylinder. 

A useful observation about the current graded UFC is that the dube algebras are isomorphic in each $\mb{g}$ sector. This implies that each defect sector is a ``copy'' of the anyons in the identity sector. Let us consider this identity sector. A subset of the irreducible central idempotents can be written down explicitly: 
\begin{align}
	\underline{a_L}=\frac{1}{\mathcal{D}^2}\sum_{x,y}\sqrt{\frac{d_xd_y}{d_a}}(R^{xa}_y)^* 
	\begin{array}{c}
\begin{tikzpicture}[scale=.1]
\def\dx{-1};
\def\ddx{1.5};
\def\dy{.5};
\def\p{5.5};
\filldraw[UFCBackground](0,0)--(20,0)--(20,20)--(0,20)--(0,0);
\draw[anyon](13,9)--(20,9) node[pos=0.55,above] {$a_\mb{1}$};
\draw[anyon](7,11)--(13,9) node[pos=0.45,below] {$y_\mb{1}$};
\draw[anyon](0,11)--(7,11) node[pos=0.55,above] {$a_\mb{1}$};
\draw[anyon](10,0)--(13,9)  node[pos=0.3,right] {$x_\mb{1}$};
\draw[anyon](7,11)--(10,20) node[pos=0.55,right] {$x_\mb{1}$};
\draw[draw=black!29,dashed](0,0)--(20,0) (20,20)--(0,20);
\filldraw[draw=black!29,fill=black!29](\p,0)--(\p+\dx,-\dy)--(\p+\dx,\dy)--(\p,0);
\filldraw[draw=black!29,fill=black!29](\p,20)--(\p+\dx,20-\dy)--(\p+\dx,20+\dy)--(\p,20);
\filldraw[draw=black!29,fill=black!29](\p+\ddx,0)--(\p+\dx+\ddx,-\dy)--(\p+\dx+\ddx,\dy)--(\p+\ddx,0);
\filldraw[draw=black!29,fill=black!29](\p+\ddx,20)--(\p+\dx+\ddx,20-\dy)--(\p+\dx+\ddx,20+\dy)--(\p+\ddx,20);
	\end{tikzpicture}
  \end{array} .
%	\mathcal{T}^x_{aya}
	\label{}
\end{align}
These idempotents are obtained from the following diagram~\cite{ocneanu1994chirality,ocneanu2001,aasen2017fermion}:
\begin{equation}
	\underline{a_L}=\frac{1}{\cal{D}^2}\sum_{x\in \cal{B}}d_x
	\raisebox{-.9cm}{
	\begin{tikzpicture}[scale=.1]
	\def\dx{-1};
	\def\ddx{1.5};
	\def\dy{.5};
	\def\p{9};
	\filldraw[UFCBackground](0,0)--(20,0)--(20,20)--(0,20)--(0,0);
	\draw[anyon](0,8)--(9, 9.8) node[pos=0.55, below] {$a_\mb{1}$};
	\draw[line width=1pt,draw=black] (11, 10.2) -- (20,12);
	\draw[anyon]  (8,0)--(12,20) node[pos=0.7, right] {$x_\mb{1}$};
	\draw[draw=black!29,dashed](0,0)--(20,0) (20,20)--(0,20);
	\filldraw[draw=black!29,fill=black!29](\p,0)--(\p+\dx,-\dy)--(\p+\dx,\dy)--(\p,0);
	\filldraw[draw=black!29,fill=black!29](\p,20)--(\p+\dx,20-\dy)--(\p+\dx,20+\dy)--(\p,20);
	\filldraw[draw=black!29,fill=black!29](\p+\ddx,0)--(\p+\dx+\ddx,-\dy)--(\p+\dx+\ddx,\dy)--(\p+\ddx,0);
	\filldraw[draw=black!29,fill=black!29](\p+\ddx,20)--(\p+\dx+\ddx,20-\dy)--(\p+\dx+\ddx,20+\dy)--(\p+\ddx,20);
		\end{tikzpicture}
	},
	\label{eqn:anyon_idem}
\end{equation}
where the crossing is resolved using the braiding in the MTC $\cal{B}$.
To prove that $\underline{a_L}\, \underline{b_L} = \delta_{ab}\underline{a_L}$, one can use the graphical calculus for braided fusion categories to simplify the diagrams. These idempotents can be identified with anyons in the chiral half of the doubled topological order.
Similarly, with the opposite crossing in Eq.~\eqref{eqn:anyon_idem} one defines $\underline{a_R}$, which is the anti-chiral half.
%Generally $\underline{a_Lb_R}$ is a degenerate block when both $a$ and $b$ are non-Abelian.

Next we consider the symmetry action on anyons. To this end,  we consider the following sector that represents a $\mb{g}$ domain wall:
\begin{align*}
\begin{array}{c}
\begin{tikzpicture}[scale=.1]
\def\dx{-1};
\def\ddx{1.5};
\def\dy{.5};
\def\p{5.5};
\filldraw[UFCBackground](0,0)--(20,0)--(20,20)--(0,20)--(0,0);
\draw[anyon](13,9)--(20,9) node[pos=0.5,above] {$a_\mb{1}$};
\draw[anyon](7,11)--(13,9) node[pos=0.25,below] {$b_{\ol{\mb{g}}}$};
\draw[anyon](0,11)--(7,11) node[pos=0.55,above] {$a_\mb{1}$};
\draw[anyon](10,0)--(13,9)  node[pos=0.3,right] {$c_\mb{g}$};
\draw[anyon](7,11)--(10,20) node[pos=0.55,right] {$c_\mb{g}$};
\draw[draw=black!29,dashed](0,0)--(20,0) (20,20)--(0,20);
\filldraw[draw=black!29,fill=black!29](\p,0)--(\p+\dx,-\dy)--(\p+\dx,\dy)--(\p,0);
\filldraw[draw=black!29,fill=black!29](\p,20)--(\p+\dx,20-\dy)--(\p+\dx,20+\dy)--(\p,20);
\filldraw[draw=black!29,fill=black!29](\p+\ddx,0)--(\p+\dx+\ddx,-\dy)--(\p+\dx+\ddx,\dy)--(\p+\ddx,0);
\filldraw[draw=black!29,fill=black!29](\p+\ddx,20)--(\p+\dx+\ddx,20-\dy)--(\p+\dx+\ddx,20+\dy)--(\p+\ddx,20);
	\end{tikzpicture}
\end{array} .
\end{align*}

We find that the dube representation of a $\mb{g}$ symmetry domain wall $B^\mb{g}_{a_L}$ is given by:
\begin{equation}
	B_{a_L}^\mb{g}=\sum_{x\in \cal{B}}\frac{d_x}{\cal{D}^2}
	\raisebox{-.9cm}{
	\begin{tikzpicture}[scale=.1]
	\def\dx{-1};
	\def\ddx{1.5};
	\def\dy{.5};
	\def\p{9};
	\filldraw[UFCBackground](0,0)--(20,0)--(20,20)--(0,20)--(0,0);
	\draw[anyon](0,8)--(9, 9.8) node[pos=0.55, below] {$a_\mb{1}$};
	\draw[line width=1pt,draw=black] (11, 10.2) -- (20,12);
	\draw[anyon]  (8,0)--(12,20) node[pos=0.7, right] {$x_\mb{g}$};
	\draw[draw=black!29,dashed](0,0)--(20,0) (20,20)--(0,20);
	\filldraw[draw=black!29,fill=black!29](\p,0)--(\p+\dx,-\dy)--(\p+\dx,\dy)--(\p,0);
	\filldraw[draw=black!29,fill=black!29](\p,20)--(\p+\dx,20-\dy)--(\p+\dx,20+\dy)--(\p,20);
	\filldraw[draw=black!29,fill=black!29](\p+\ddx,0)--(\p+\dx+\ddx,-\dy)--(\p+\dx+\ddx,\dy)--(\p+\ddx,0);
	\filldraw[draw=black!29,fill=black!29](\p+\ddx,20)--(\p+\dx+\ddx,20-\dy)--(\p+\dx+\ddx,20+\dy)--(\p+\ddx,20);
		\end{tikzpicture}
	} .
	\label{}
\end{equation}
The crossing is again resolved using the diagrammatic rules for braiding in the MTC $\cal{B}$, ignoring the $\mb{g}$ label (i.e. $x_\mb{g}$ is just treated as $x$). It then obviously satisfies
\begin{equation}
	B^\mb{g}_{a_L}\underline{a_L}=\underline{a_L}B^\mb{g}_{a_L},
	\label{}
\end{equation}
which again shows that no anyons are permuted.

Furthermore, expanding $B^\mb{g}_{a_L}$ in the dube basis, a direct computation yields
\begin{equation}
	B^\mb{g}_{a_L}B^\mb{h}_{a_L}= \eta_{a}^{-1}(\mb{g,h})B^{\mb{gh}}_{a_L}.
	\label{}
\end{equation}
This identity shows that $a_L$ transforms as a projective representation under $G$, with the factor set given by $\eta_a^{-1}$.

\bibliography{cft}

\end{document}